\documentclass[reqno,11pt]{amsart}
\usepackage{hyperref}
\usepackage{graphicx}
\usepackage{slashed}
\usepackage{amscd}
\usepackage{tensor}
\usepackage{amssymb}
\usepackage{esint}
\usepackage{enumitem}
\usepackage{accents}
\usepackage[dvipsnames]{xcolor}
\usepackage[utf8]{inputenc}
\usepackage[off]{auto-pst-pdf}
\usepackage{pst-grad}
\usepackage{pst-plot}
\usepackage[mathscr]{eucal}
\usepackage{tagging}
\droptag{NotReady}

\newcommand\Johannes[1]{}
\renewcommand\Johannes[1]{\marginpar {\flushleft\sffamily\footnotesize {\textcolor{CornflowerBlue}{Johannes}}: #1}}

\numberwithin{equation}{section}
\allowdisplaybreaks[1]

\newtheorem{Def}{Definition}[section]

\newtheorem{Lemma}[Def]{Lemma}
\newtheorem{Remark}[Def]{Remark}

\newtheorem{Example}[Def]{Example}

\newenvironment{ExampleS}{\begin{small}\begin{Example}}{\QEDrem\end{Example}\end{small}}

\newtheorem{Convention}[Def]{Convention}

\newtheorem{PA}[Def]{Phenomenological Axiom}

\newcommand{\ch}[1]{#1}

\newcommand{\Thanks}{\vspace*{.5em} \noindent \thanks}
\newcommand{\beq}{\begin{equation}}
\newcommand{\eeq}{\end{equation}}
\newcommand{\Proof}{\begin{proof}}
\newcommand{\QED}{\end{proof} \noindent}
\newcommand{\QEDrem}{\ \hfill $\Diamond$}

\newcommand{\la}{\langle}
\newcommand{\ra}{\rangle}

\newcommand{\R}{\mathbb{R}}

\newcommand{\Z}{\mathbb{Z}}
\newcommand{\N}{\mathbb{N}}

\DeclareMathOperator{\Tr}{Tr}
\DeclareMathOperator{\Pa}{Pa}
\DeclareMathOperator{\id}{id}
\DeclareMathOperator{\Aut}{Aut}

\DeclareMathOperator{\fe}{fe}
\DeclareMathOperator{\efe}{efe}

\DeclareMathOperator{\pcl}{pcl}

\DeclareMathOperator{\Iso}{Iso}

\renewcommand{\L}{{\mathcal{L}}}

\newcommand{\G}{{\mathcal{G}}}
\newcommand{\bars}{{\bar{s}}}

\renewcommand{\H}{\mathscr{H}}

\newcommand{\K}{{\mathcal{K}}}
\newcommand{\Autk}{\Aut(\E)^n}

\newcommand{\sigmaP}{\tilde \sigma}
\newcommand{\D}{\mathscr{D}}

\newcommand{\I}{{\mathcal{I}}}

\DeclareFontFamily{OT1}{rsfso}{}
\DeclareFontShape{OT1}{rsfso}{m}{n}{ <-7> rsfso5 <7-10> rsfso7 <10-> rsfso10}{}
\DeclareMathAlphabet{\mycal}{OT1}{rsfso}{m}{n}

\newcommand{\itemD}{\item[{\raisebox{0.125em}{\tiny $\blacktriangleright$}}]}

\newcommand{\bigslant}[2]{{\raisebox{.2em}{$#1$}\left/\raisebox{-.2em}{$#2$}\right.}}	
\newcommand{\bei}{\begin{itemize}[label=-,topsep=0pt]}
\newcommand{\eni}{\end{itemize}}

\newcommand{\E}{E} 
\newcommand{\e}{e} 
\renewcommand{\P}{P} 

\definecolor{darkblue}{RGB}{0,91,163}

\newcommand{\itsl}{\item[-]}

\title[Mathematical Models of Consciousness]{Mathematical Models of Consciousness}

\author[J. Kleiner]{Johannes Kleiner$^\dagger$}
\address{Munich Center for Mathematical Philosophy\\Ludwig Maximilian University of Munich\\ Geschwister-Scholl-Platz 1 \\80539 Munich\\ Germany}

\begin{document}
\maketitle
\ \\[-3.4em]
\begin{center}\small
$^\dagger$Munich Center for Mathematical Philosophy\\Ludwig Maximilian University of Munich\\[1.6em]
\end{center}

\begin{abstract}
In recent years, promising mathematical models have been suggested which aim to describe conscious experience and its relation to the physical domain. Whereas the axioms and metaphysical ideas of these theories have been carefully motivated, their mathematical formalism has not.
In this article we aim to remedy this situation.  We give an account of what warrants mathematical representation of phenomenal experience, derive a general mathematical framework which takes into account consciousness' epistemic context and study which mathematical structures some of the key characteristics of conscious experience imply, showing precisely where mathematical approaches allow to go beyond what the standard methodology can do. The result is a general mathematical framework for models of consciousness that can be employed in the theory-building process.
\end{abstract}\bigskip

\begin{center}
\end{center}\medskip

\tableofcontents

\section{Introduction}\label{SecIntrod}
Conscious experience and its relation to the physical domain has been studied by philosophers, theologians and scientists over many centuries \cite{companion.2014.consciousnessmodernscientificstudyof}. In the previous three decades, there has been a resurgence of scientific investigations. Groundbreaking developments in neuroscience, cognitive psychology and analytic philosophy lead to the emergence of a dedicated \emph{science of cosnciousness}, whose aim is to develop a scientific account of conscious experience and its relation to the physical domain (e.g. brain processes).

A model of consciousness is a hypothetical theory about how conscious experience and the physical domain relate \cite{Seth.2007}. Examples
include Global Workspace Theories \cite{DehaeneKerszbergChangeux.1998,baars2005global}, Multiple Draft Theory~\cite{dennett1993consciousness}, Higher Order Thought Theories~\cite{sep-consciousness-higher} or Integrated Information Theory~\cite{IIT30}, among many others.
Models of consciousness complement metaphysical theories of consciousness, such as the various forms of functionalism, identity theories, interactive dualisms or neutral monisms. These theories are concerned primarily with ontological questions and address the general type of relation between consciousness and the physical domain. 

\addtocontents{toc}{\protect\setcounter{tocdepth}{1}}
\subsection*{The rising importance of mathematics in consciousness studies}

In recent years, many models of consciousness have been proposed which are mathematical in nature. Primary examples in neuroscience are the recent versions
of Integrated Information Theory~\cite{IIT30,Mayner.2018,haun2019does}, which aim to determine the quality and quantity of a system's conscious experience using a
complex mathematical algorithm~\ch{\cite{kleiner2020mathematical,tull2020integrated}}, or Predictive Processing Theory~\cite{ppp.2017}, which can be interpreted as specifying the content of conscious experience of a system using an advanced minimization principle~\cite{dolkega2020fame}. But promising models have been proposed by other disciplines as well, including philosophy~\cite{chalmers2014consciousness}, physics~\cite{kent2018quanta,kent2019toy}, mathematics~\cite{penrose1994shadows,kremnizer2015integrated,Mason.2016} or psychology \cite{HoffmanPrakash.2014}, all based on various 
different metaphysical ideas about consciousness.

These developments point at an increasing relevance of mathematical tools and methods in the scientific study of consciousness, much like in other scientific disciplines throughout the last century, with promising new insights on the horizon. However, mathematization on its own does not have unique scientific merit. Valuable progress can only be made if the mathematization is based on and integrates previous theoretical, empirical and conceptual work.

\subsection*{What makes consciousness a problem}
Consciousness is a phenomenon unlike any other studied by natural science. It is unique as an object of investigation both in its characteristic features and in its epistemic context. This is true in particular for its most relevant and mysterious connotation, which much of the science of consciousness and also this article is concerned with, namely phenomenal consciousness. Phenomenal consciousness  refers to the way in which the world \emph{appears} to us, i.e. the way in which we experience the world. This can roughly be paraphrased as ``pure subjective experience''~\cite{metzinger1995problem}. 

Much of philosophy of mind is concerned with analysing in detail just what the characteristic features of phenomenal consciousness are and how precisely they relate to metaphysical ideas and efforts of scientific investigation. The arguably most crucial features attributed to phenomenal experience are its essential \emph{subjective nature}, which sometimes is taken to mean that phenomenal consciousness embodies a particular point of view~\cite{Nagel.1974}, but also that some of its parts or properties seem \emph{ineffable}~\cite{lewis1929mind}, \emph{private}, or \emph{unavailable to cognitive and linguistic processing or communication}~\cite{metzinger2007grundkurs}.

Basic properties or simple constituents of phenomenal consciousness are called \emph{qualia}, but this term is being used with many different connotations to date. Qualia are variously claimed to be \emph{intrinsic} and \emph{non-relational} or to have a \emph{qualitative and non-quantifiable} nature. Phenomenal consciousness is also claimed to be \emph{directly or immediately apprehensible}, to be \emph{transparent} in the sense that it appears as if we are in direct contact with the content of our conscious experience or to be \emph{homogeneous} \cite{metzinger1995problem}. 
Various different \ch{connotations} of all of these notions exist, and different philosophers endorse various combinations thereof.

Complementing these characteristic features of consciousness is its unique epistemic context,
which comes about from the fact that phenomenal consciousness per se is accessible only to the experiencing system itself. Thus in any scientific approach there are two fundamentally different methodological approaches that allow \ch{one} to gather information, a first person perspective and a third person perspective. 
This is referred to as the \emph{epistemic asymmetry} of consciousness~\cite{metzinger1995conscious}.

\subsection*{The need for a mathematical foundation}

Any scientific analysis which strives to address and explain phenomenal consciousness needs to take these features of consciousness into account, at the very least as providing epistemic restrictions which constrain and shape the empirical access to conscious experience. Failing to do so at all amounts to ignoring what makes consciousness a problem in the first place, which no serious scientific investigation can afford.

To date, almost none of the existing formal models take any of these properties of consciousness into account. While mathematical structures are quickly associated with terms like  `qualia', `subjective experience' or `act of consciousness', contemporary models fall short of actually considering the conceptual meaning of these philosophical concepts. 

What is necessary to mend this is a thorough foundation of mathematical models of consciousness that analyses which implications the various characteristics of conscious experience have on the mathematical structure of these models and which provides a precise account of how the concepts developed in philosophy of mind relate to the mathematical structure of models of consciousness.

The goal of the work presented in this article is to provide this foundation for the case of {ineffability}, {privateness} and {cognitive, linguistic and communicative unavailability}.

\subsection*{A framework for formulating models of consciousness}

The result of this work is a general mathematical \emph{framework} in which models of consciousness can be formulated. Much like Lagrangian mechanics in theoretical physics, it does not provide any particular law or equation which constitutes a model of consciousness, but rather a general formal machinery. What this machinery achieves is to properly take into account that conscious experience has ineffable, private or inaccessible aspects and that it exhibits an epistemic asymmetry. This framework provides a first mathematical foundation for models of consciousness, and needs to be expanded in future work to take other key characteristics into account. 

Crucially, the framework is independent of whether one considers any of these characteristic features to be ontological in origin or simply due to a system's particular design or cognitive functions~\cite{dennett1993consciousness}. What matters, from the perspective of this framework, are only the epistemic restrictions that arise from these features of conscious experience, i.e. that access to some parts of conscious experience is limited by consciousness' subjective nature and by \ch{ineffability}, {privateness} and {inaccessibility}  in any type of experimental situation.

Thus this framework can be dubbed \emph{operational}. Much like Quantum Theory in its conventional formulation, it takes as its starting point the prototypical experimental situation in which a theory (of consciousness) is being tested, used or inferred, and then adds the particular epistemic context of consciousness, so as to arrive at a general operational description.
Great care has been taken to keep the mathematical structure of this formalism as general as possible, and to provide operational justifications of all essential definitions, so as to ensure that the framework is \emph{compatible} with all types of mathematical structure one would want to use in modelling consciousness, including category theory~\cite{tull2020integrated,tsuchiya2016using}, information theory~\cite{Tononi.2008} or complex system approaches~\cite{atmanspacher2016macrostates}, among many others. 



\subsection*{An axiomatic conceptual underpinning}
In order to translate any concept into formalism, the concept itself needs to be rigorously defined. To date, neither the term qualia, nor the concepts of \ch{ineffability}, {privateness} and {inaccessibility} are defined in a rigorous enough manner to warrant thorough formalization.

Thus in order to achieve our goal, it was necessary to represent the underlying philosophical concepts in an axiomatic form that is suitable for formalization. Since this whole programme is operational in nature, it suffices in fact to provide an axiomatic definition of the \emph{operational consequences} of these characteristic features of consciousness.

For the case of \ch{ineffability}, privateness and inaccessibility this is possible at once by introducing the concept of \emph{non-collatability}: A part, property or feature of conscious experience is non-collatable if there do not exist any reasonable means to identify this part, property or feature over several experiencing subjects in an experimental trail.
Non-collatability is entailed by \ch{ineffability}, privateness and inaccessibility and arguably also by consciousness' subjective nature.
As shown in detail throughout this work, it is precisely non-collatability which generates much of the epistemic difficulty in investigating conscious experience, and which has substantial consequences for any empirically adequate model of consciousness.

The conceptual definitions we have derived in order to found the mathematical structure of models of consciousness give rise to an axiomatic \emph{grounding} of the scientific study of consciousness that is an alternative to and further development of the grounding that derives from David Chalmers' work. 
While our grounding was primarily intended to constitute an intermediate construction which links philosophical concepts and mathematical formalism, it may also have some conceptual worth in its own, providing an interim way to conceive of the task and methodology of the scientific study of consciousness from a more formal perspective.

\subsection*{A new way of consciousness science}
In summary, this article can be understood as taking seriously a new way of doing consciousness science that has been pioneered in
eminent works such as \cite{IIT30}, \cite{HoffmanPrakash.2014} or \cite{Resende.Talk.2018}.
Its central idea is to represent phenomenal consciousness in terms mathematical spaces, and to use these spaces to build theories of how conscious experience might relate to the physical domain. This \ch{facilitates} a much richer and refined way of
addressing \ch{conscious experience} and offers promising tools to resolve of some of the key issues that permeate contemporary consciousness studies.

What this article adds to this new methodology is the requirement that the essential features of conscious experience, studied in detail by philosophy of mind, are taken into account when building this mathematical representation. Doing so requires an account of how precisely mathematical spaces can be grounded in the phenomenology of experience and of which mathematical implication consciousness' fundamental epistemological context has. Answers to all of these questions are proposed here. The hope is that these considerations might provide a useful basis for further development of formal models of consciousness.


\subsection*{The structure of this article}
This article is structured as follows. In order to make it accessible to readers without formal background, we first summarize the results in Section~\ref{sec:summary}, keeping mathematical details to a minimum. In this section, we also explain in detail the rationale and motivation of this work.

All subsequent sections aim for a concise presentation of definitions, explanations and examples. In Section~\ref{NewGrounding}, we give the conceptual definitions on which our framework rests, making as few assumptions as necessary. This gives rise to an axiomatic grounding of the scientific study of consciousness. In Section~\ref{ExplGap}, we show that there is an explanatory gap between qualia as defined here and natural science. 
Section~\ref{SecMathStrMoc} is devoted to deriving a general mathematical framework for formal models of consciousness, making use of a minimal set of ingredients of any formal theory and of consciousness' epistemic asymmetry. In Section~\ref{PhenGrMethod}, finally, we show how consciousness' characteristic features can be taken into account. We conclude this paper with a brief remark on a metaphysical question in Section~\ref{ClosureOfPhys} and various examples in Section~\ref{Examples}.

In Appendix~\ref{ChalmersGrounding}, we review the grounding of the scientific study of consciousness that derives from David Chalmers' work in~\cite{Chalmers.1996,Chalmers.2010}, emphasizing the logical relations among its parts. In Appendix~\ref{APIssuesChalmers}, we discuss problems that arise if one attempts to apply this grounding in a model-building process.

\addtocontents{toc}{\protect\setcounter{tocdepth}{2}}
\section{Summary of Results}\label{sec:summary}

Any research activity directed at conscious experience presupposes a conception of the phenomenon that is to be studied and a conception of a methodology that is suitable to do so. We call this a grounding of the scientific study of consciousness.

\begin{Def}\em\label{DefG}
A  \emph{grounding of the scientific study of consciousness} contains at least
\bei
\itsl an explicit definition of what is to be studied.
\itsl an explicit outline of the methodology.\footnote{Here ``methodology'' refers to ``a collection of methods, practices, procedures and rules used by those who work in some field''~\cite{Wiktionary.methodology}, ``a system of methods used in a particular area of study or activity''~\cite{Oxfordic.methodology}. In particular, the methodology includes the specification of what constitutes an {\em experiment}. The term `grounding' is one of several translations of the German word ``Grundlegung''.}
\eni
\end{Def}

Much of the research devoted to consciousness in the previous two decades has been guided by a grounding that derives from David Chalmers' work~\cite{Chalmers.1996}. This grounding has played a pivotal role in the creation and consolidation of the field. However, it also exhibits several severe problems when being applied (Appendix~\ref{APIssuesChalmers}). In the first part of this paper, we introduce an alternative to Chalmers' grounding of the scientific study of consciousness. This alternative is built on a \emph{thoroughly operational perspective}, which means that we define all notions relative to prototypical experimental investigations. 

Any experimental situation devoted to study conscious experience presupposes a preliminary choice of organisms that are considered to be conscious, and whose conscious experience and physical state is probed during the experiment in order to gain information about how consciousness relates to the physical domain. We denote any such class of organisms by $\mathcal C$ and call them \emph{experiencing subjects}. Taking the operational perspective seriously, we consider $\mathcal C$ to be a primitive notion. While it may be guided by theoretical insights and changed over the course of time, at any particular time a class $\mathcal C$ provides the basis for both inference and tests of theories about consciousness.

Having chosen our primitive notion, we can define experience relative to it. A promising choice is to use somewhat phenomenological terminology in defining the term conscious experience, referring to the \emph{totality} of how experience `reveals itself' to an experiencing subject, how the experiencing subject finds itself experiencing, or how the `the world’ appears to it. 
While this is what we have in mind, we have opted for more approachable terminology and define the term \emph{conscious experience} to denote totality of impressions, feelings, thoughts, perceptions, etc. which an experiencing subject lives through at a particular instant of time (Definition~\ref{DefExperience}). Experience so defined has various different \emph{aspects}, where we define the term `aspect' to be a placeholder for any conception like `part', `property' or `element of' (Definition~\ref{DefExp}).

\ch{The key notion on which our grounding is built is that of non-collatability. An aspect of experience is \emph{non-collatable} if there is no reasonable method to identify whether two or more experiencing subjects in an experiment experience this very aspect of experience. In other words, if the identity of this aspect over several different experiencing subjects in $\mathcal C$ cannot be determined (Definition~\ref{Noncollatable}).}

The distinction between collatable and non-collatable aspects of experience is what replaces the distinction between phenomenological and psychological concepts of mind in~\cite{Chalmers.1996}. Whereas the latter distinction is defined in terms causal roles and spatiotemporal structure (cf. Appendix~\ref{ChalmersGrounding}), our distinction is defined axiomatically in terms of phenomenological or operational notions.

What is crucial is that non-collatability is implied by various essential characteristic features of conscious experience.
E.g., any aspect which appears to be \emph{ineffable} (i.e. which is \emph{experienced} as ineffable) is also non-collatable in the above sense. The same is true for aspects of experience which are experienced as \emph{private} or which are \emph{not available
to cognitive or linguistic processing}. All of these characteristic features destroy the possibility to identify an aspect under consideration over several experiencing subjects. Non-collatability is a necessary operational consequence of all of these characteristic features.

The same may be true of subjectivity of conscious experience, if taken to warrant the claim that ``there are facts that do not consist in the truth of propositions expressible in a human language''~\cite[p.\,441]{Nagel.1974}. In fact, one of the main claims in~\cite{Nagel.1974} is that there is at present no conception that allows \ch{one} to establish the identity of a `what it is like' aspect of experience with a physical state. Our starting point, non-collatability, is closely related to this claim and may even be implied by it in reasonable cases.

Building on non-collatability, we define qualia as follows.\footnote{
Note that the numbering of this and all following definitions is chosen according to the main body of this article.}

\bigskip\noindent {\bf Definition~\ref{DefQualia}}
We define the term {\em qualia} to refer to all non-collatable aspects of experience of an experiencing subject within the class $\mathcal C$.
\bigskip

This definition is warranted since it includes the paradigmatic examples of what qualia are claimed to be (Example~\ref{ExColourQualia}), as well as aspects of experience referenced by the Nagelian `what it is like' conception (Example~\ref{ExWhatIsItLike}). It is furthermore axiomatic and replaces the concept of phenomenal consciousness as defined in Chalmers' grounding.\footnote{We remark that for this and all other definitions, it does not matter whether non-collatability or any of the characteristic features which imply it are considered to be fundamental or merely the result of a system's architecture. All that matters is that experience appears as such.}

The aspects of experience which satisfy Definition~\ref{DefQualia} are of special interest because the non-collatability
induces a fundamental difficulty in any scientific approach: It implies that these aspects cannot be referenced intersubjectively,
which in turn implies that they cannot be referenced in a scientific model or empirical analysis. There is a fundamental explanatory gap (Section~\ref{ExplGap}).
The goal of this paper, when put in these terms, is to develop a mathematical framework that allows \ch{us} to address both collatable and non-collatable aspects of experience, providing a formal methodology suitable to address this explanatory gap.\medskip

Next, we make use of the central idea that underlies many contemporary mathematical models of consciousness: To represent phenomenal consciousness as a mathematical space. In order to provide an accurate method to do so, we make use of two phenomenological \ch{axioms}.

First, we make use of the fact that both collatable and non-collatable aspects of experience can be recognized to a certain extent (Phenomenological \ch{Axiom}~\ref{PFRecognize}). Another way to say this is that aspects of experience may be experienced as identical. Following our operational perspective, this warrants the introduction of \emph{labels} for both qualia and collatable aspects of experience, i.e. names relative to an experiencing subject. 

Second, we make use of the fact that there are collatable relations between aspects of experience (Phenomenological \ch{Axiom}~\ref{PFRel}). This might be considered obvious in the case of collatable aspects of experience. With respect to non-collatable aspects of experience, it corresponds to the observation that ``structural features of perception might be more accessible to objective description, even though something would be left out''~\cite[p.\,449]{Nagel.1974},
or that ``even if experiences are in some sense `ineffable,' relations between experiences are not; we have no trouble discussing these relations, whether they be relations of similarity and difference, geometric relations, relations of intensity, and so on. As \ch{Schlick~\cite{schlick1969form} pointed out, the \emph{form} of experience seems to be straightforwardly communicable, even if the \emph{content} (intrinsic quality) is not}''~\cite[p.\,224]{Chalmers.1996}.

Together, these two phenomenological observations allow us to define a mathematical space that \ch{\emph{represents}} conscious experience,
which we call \emph{experience space} $\E$. \ch{The elements of this space are not experiences themselves but \emph{labels} that an experiencing subject may give for his/her aspects of experience, and the mathematical structure on this set of labels is induced by the collatable relations between aspects of experience.
To conceive of $E$ as space of labels, rather than as a space of experiences, is of advantage because it prevents from the very beginning any implicit assumption of well-defined reference to aspects of experience. In contrast, working with a space whose elements are intended to express experiences themselves requires the introduction of a map which describes how these experiences can be inferred from reports (labels), e.g. as in~\cite{kleiner2020falsification}.
The details of our introduction of experience space are explained in Section~\ref{SecFormal}, and various examples are given in Section~\ref{SecExamplesBasic}.}

We remark that whereas our constructions are guided by conceiving of labels as something that an experiencing subject can express, which requires $\mathcal C$ to comprise humans, this is not necessary. This is to because the various principles that are used in experiments to date to infer the state of consciousness of some subject (e.g. button presses or behavioural indicators) are, in our terminology, in fact means to infer labels of aspects of experience. Whether a label is a recorded word or some other type of report, such as a particular movement, does not matter for our purposes.
What is crucial about the terminology of labels is that one avoids from the very beginning any implicit assumption that there is an empirically well-defined method to refer to qualia of an experiencing subject.
\medskip

Non-collatability implies limitations on how aspects of experience can be referenced in a theory or empirical investigation. Whereas labels of collatable aspects of experience can be synchronized over all experiencing subjects in the class $\mathcal C$ (because collatability holds iff there exist means to identify), labels of qualia cannot. In virtue of non-collatability, the definition of qualia implies that any label which one experiencing subject uses to denote a quale may denote another quale in a different experiencing subject. Any scientific investigation which aims to address qualia needs to take the resulting ambiguity into account.
Ignoring it will lead to errors, such as the study of the wrong ``information pathway'' or confounding neural correlates of external signals with neural correlates of qualia.

In the next step of our construction, we quantify this ambiguity precisely. To this end, we make use of the mathematical representation $\E$ of experience constructed previously. As we explained in detail in Section~\ref{SecQualiaRef}, the conceptual and mathematical definitions imply that the ambiguity of any reference to aspects of experience can be stated concisely in terms of the automorphism group $\Aut(\E)$ of $\E$, i.e. the group of all transformations of $\E$ which change labels in such a way that the mathematical structure of $\E$ is left invariant.

We find that any statement about conscious experience that uses an individual label~$e$ could, in light of non-collatability, have equally well be formulated in terms of any label $e'$ that is part of a subset $[e]$ of labels. This subset $[e]$ is called the equivalence class of the label $e$ with respect to the automorphism group $\Aut(\E)$.

The crucial insight here is that these equivalence classes describe what is intersubjectively accessible or, in our terminology, empirically well-defined. Taken together, these equivalence classes describe what is amenable to the usual scientific methodology. Depending on the mathematical structure of $\E$, this may well include some of the non-collatable aspects of experience.

This second step in our constructions may be sufficient for many investigations once experimental tools become advanced enough to proceed to the study of individual aspects of experience. It \ch{enables} experimentalists to use structural features of phenomenal experience, as represented in $\E$, to push the boundary of ineffability and all the other characteristic features that imply non-collatability back a little bit.
However, as long as there are equivalence classes $[e]$ which contain more than one label, there are questions that evade the reach of the standard scientific methodology: Why the subject had one of the corresponding experience rather than the other. Even \ch{though} this cannot be expressed \ch{by} intersubjective means, there is a fact to the matter for the experiencing subject, and hence a priori an open scientific question. The goal of the remaining part of the article is to develop tools that allow us to address this open question.\medskip

To develop these tools, we have to go beyond the mathematical representation of experience, and in fact consider formal \ch{hypotheses} about how conscious experience relates to the physical domain, i.e. formal models of consciousness. In order to remain as general as possible, as a first step in answering this question, we ask what the most general mathematical structure is that a model of consciousness needs to address.

In order to answer this question, we first give an account of the minimally sufficient formal structure of any scientific theory (Section~\ref{DefTheories}).
A theory needs to specify some dynamical variables $d$ that describe what the theory intends to address, may contain some formal background structure, needs to have a parameter such as time that \ch{facilitates description} of variations of the dynamical variables and, finally, needs to contain some laws that pick out some variations of $d$ from all possible variations.

In order to further fix the dynamical variables $d$, we make use of the epistemic asymmetry of conscious experience (Section~\ref{PreMods}).
The epistemic asymmetry states that there are two fundamentally different ways of gathering knowledge about conscious experience, the first-person perspective and the third-person perspective. Thus there are two \emph{epistemically} different notions of state in any experimental situation, one that corresponds to first-person access, and one that corresponds to third-person access. While any metaphysical theory of consciousness can ignore one of these states,
a scientific model of consciousness cannot. The difference between a coherent idea and a scientific model of consciousness is precisely that the latter addresses both types of states, while the former need not.

Since the states that are accessible in the third-person perspective are in fact physical states (neural states, brain states, or similar)
and the states that are accessible in the first-person perspective are aspects of experience (with `aspect' suitably defined, cf. above), this implies that the dynamical variables of a formal model of consciousness are in fact a \ch{subset of}
$$
	d= \E \times \P \:,
$$
where~$\E$ denotes the mathematical representation of conscious experience we have introduced above and where~$\P$
denotes the state space of some physical theory $T_P$.

Combining the above gives a general framework in which models of consciousness (Definition~\ref{DefPreModel}) can be formulated. It provides a reference to which models of consciousness \emph{need to refer} in light of consciousness' epistemic context, 
independently of how they are primarily defined and independently of which ontological ideas they express.
\medskip

This general framework finally puts us into a position to investigate the implications of non-collatability in Section~\ref{PhenGrMethod}.
First, in Section~\ref{SecNecWellDef}, we prove that in light of non-collatability, models of consciousness are only well-defined if they carry a particular symmetry. 
This is comparable to physical theories. Much like general relativity carries a particular symmetry that ensures that the theory is well-defined with respect to changes of coordinates, our results show that models of consciousness need to carry a particular symmetry that ensures that they are well-defined with respect to changes of labels. The changes of labels in question are precisely those transformation which keep the equivalence classes $[e]$ introduced above constant, but transform individual members of these equivalence classes. The corresponding symmetry group is the automorphism group $\Aut(\E)$ introduced above.

What is crucial in our results of Section~\ref{SecNecWellDef} is that the symmetry required to exist is not fixed uniquely. There is a freedom in its form which 
depends on the laws of a model of consciousness. This freedom describes how the transformations of labels relate to transformations of physical states. Sections~\ref{SecDefMoc} and~\ref{Comparison} are devoted to proving that this remaining freedom is what allows formal models of consciousness to go beyond what the standard methodology can do.

In a nutshell, this is so because the standard methodology can only utilize intersubjectively well-defined references. Mathematically, this means that the standard methodology can only reference aspects of experience once it has imposed the group \ch{$\Aut(\E)$} in order to construct equivalence classes $[e]$. Formal models of consciousness, on the other hand, allow one to reverse this order. They allow one to relate individual elements of $\E$ and $\P$ \emph{prior to} imposing the symmetry which ensures well-definedness.

Since all our arguments, proofs and derivations hold true also in the limiting cases where all aspects of experience are either collatable or non-collatable, we summarize all insights in a concise definition of what a model of consciousness is (Definition~\ref{DefModel}). This is the main result of our project. 

\begin{flushright}\em\small
``Many scientific discoveries have been delayed over the\\
 centuries for the lack of a mathematical language\\
that can amplify ideas and let scientists\\
communicate results.''~\cite[p.\,427]{Pearl.2009}
\end{flushright}

\section{Basic Definitions}\label{NewGrounding}

In this section, we provide the basic definitions that underlie our constructions. In Section~\ref{PhenomenologicalGrounding}, we specify the notion of experience we consider, introduce some fundamental terminology and give a definition of qualia in these terms. Subsequently, in Section~\ref{SecFormal}, we discuss the mathematical representation of experience. In Section~\ref{SecQualiaRef}, we explain the implications of the defining characteristic of qualia for any reference to consciousness in an experiment or theory. As mentioned above, altogether this can be taken to provide a grounding of the scientific study of consciousness, and Section~\ref{SecPhenGr} is devoted to summarize the resulting picture. In Section~\ref{SecExamplesBasic}, finally, we give several examples for the mathematical structure introduced in Sections~\ref{SecFormal} and~\ref{SecQualiaRef}.

\subsection{Conscious Experience and Qualia}\label{PhenomenologicalGrounding}
The starting point of every {\em scientific} activity related to consciousness is a preliminary choice of a class $\mathcal C$ of experiencing subjects
that are available for experimental investigations and which are targeted by theoretical models.
The object of investigation of any empirical study, and what informs any model-building process, is experience of these experiencing subjects in the following sense.

\begin{Def}\em\label{DefExperience}
We use the term `{\em conscious experience}' (`{\em experience}' for short) to denote the {\em totality} of impressions, feelings, thoughts, perceptions, etc. which an experiencing subject lives through at a particular instant of time.\footnote{No special focus on subjectivity is intended when using the term `experiencing subject'. Alternatively, one could use the term `experiencer'. We also remark that the meaning of `instant' is to be fixed during the model-building process. It could refer to physical just as well as to experiential instants of time.}
\end{Def}

The {\em general idea} underlying {\em any} conception of the scientific study of consciousness is to study experience and its relation to the physical domain by scientific means. Mostly, some part or feature of experience is under consideration. In order to emphasise that this part or feature may not be strictly separable from other parts of features, we use the term `aspects of experience':

\begin{Def}\em \label{DefExp}
{\em Aspects of experience} denote specific or general features, parts, properties or elements of a particular experience or of a set of  experiences.
\end{Def}

\noindent According to this definition, `aspect' is thus merely a placeholder for `feature', `part', `property' or `element of'. Which of these notions is relevant is part of the specification of a model of consciousness.

\begin{ExampleS}\em
Aspects of experience range from individual visual, auditory or tactile experiences to general characteristics, such as the experience of a first person perspective, the unity of the conscious scene~\cite{Seth.2007}, or the structure and composition of experience~\cite{IIT30}.
~\end{ExampleS}

\noindent A priori, every experiencing subject only has access to his/her own experience. However, systematic investigations of which aspects of experience
are invariant over a large class of experiencing subjects are possible and have been carried out as part of 
the philosophical discipline of phenomenology.

\begin{Def}\em \label{DefPA}
A {\em phenomenological \ch{axiom}}
is a statement about aspects of experience which holds for all experiencing subjects in a class~$\mathcal C$.%
\footnote{\label{FootnoteClassC}The restriction to a class~$\mathcal C$ of experiencing subjects is necessary because a phenomenological analysis of invariants of experience is always restricted to experiencing subjects which are similar in some respects:
``[O]ne person can know or say of another what the quality of the other's experience is. [However, this] ascription of experience is possible only for someone sufficiently similar to the object of ascription to be able to adopt his point of view''~\cite[p.\,442]{Nagel.1974}.
However, the choice of class~$\mathcal C$ is not a constraint for models of consciousness, but rather a~{\em starting point}, i.e. a preliminary choice which informs the model-building process. Models may eventually allow one to determine which organisms experience.
We note also that the name `phenomenological \ch{axiom}' is a tribute to phenomenology rather than an attempt to condense the phenomenological method into a simple definition.}
\end{Def}

\noindent Phenomenological \ch{axioms} serve as a starting point for any investigation in the scientific study of consciousness. In empirical studies, they are what can be correlated with physical states, e.g. to construct neural correlates of consciousness. When building models of consciousness, they are what informs the choice of mathematical structure.
In simple terms, phenomenological \ch{axioms} are statements about how experiencing subjects find themselves experiencing, or how `the world' appears to them.
One could also say that they express invariant facts of `what experience is like' or of how experience `reveals itself'.

\smallskip

Next, we make use of three basic phenomenological \ch{axioms} in order to examine in more detail which methodology may be used to study aspects of experience and their relation to the physical domain. In preparation, we define the concept of non-collatability.

\begin{Def}\em \label{Noncollatable}
\ch{An aspect of experience is \emph{non-collatable} iff there does not exist a reasonable method to establish its identity over different experiencing subjects in the class~$\mathcal C$.\footnote{
\ch{I.e., an aspect experienced by subject $\mathcal S_1$ is non-collatable iff there is a different experiencing subject $\mathcal S_2$ such that there is no reasonable method to determine which aspect $e'$ of $\mathcal S_2$ the aspect $e$ of $\mathcal S_1$ is identical with. Put yet in different terms, this is the case if there is no  mapping from $e$ to the aspects of experiences of other experiencing subjects that can reasonably be interpreted as establishing identity of aspects.}
}}

\end{Def}

\ch{Non-collatability of an aspect of experience can be determined operationally in any experimental situation. Whenever there is no reasonable method to identify whether
two experiencing subjects in an experiment experience the same aspect of experience, the aspect in question is non-collatable. However, the concept can also be applied in a more fundamental context, as part of a phenomenological axiom about how subjects experience the world. Several examples are given below.}

\begin{PA}\em \label{PFcollatable}
Aspects of experience can be divided into two classes:
\begin{enumerate}[label=\alph*),topsep=0pt]
\item Aspects of experience which are \emph{non-collatable}.
\item Aspects of experience which are \emph{collatable}.
\end{enumerate}
\end{PA}

The former class includes aspects of experience which are experienced as \emph{ineffable}~\cite{lewis1929mind}, \emph{private}, or found to be \emph{cognitively, linguistically and communicatively inaccessible}~\cite{metzinger2007grundkurs}. But they also include those which are referred to as having a \emph{subjective character}~\cite[p.\,437]{Nagel.1974} or connected to a \emph{particular point of view}~\cite[p.\,441]{Nagel.1974}. Non-collatability is implied by, and hence a necessary condition of, all of these characteristic features of experience. 
The latter class include those aspects which are experienced as accessible also \emph{from other points of view}~\cite[p.\,443]{Nagel.1974} or as having an \emph{objective nature}~\cite[p.\,443]{Nagel.1974}.%
\footnote{When being presented with Phenomenological \ch{Axiom}~\ref{PFcollatable}, scientists usually tend to think about how this can be derived from a theory of language. In our opinion, the more important task is to ground the underlying distinction in a thorough phenomenological analysis. We also remark that all formal constructions in this article are compatible with either of the classes in Phenomenological \ch{Axiom}~\ref{PFcollatable} being empty, even though this is most likely not the case.}
 
\begin{ExampleS}\em
Consider, as a first example, experiences of awe. Subjects may report that they have an experience of awe, and even give labels to various different such experiences, but there is, at present, no methodological procedure to establish whether any two experiences of awe of two different subjects are the same or not. 

Some~\cite{churchland1981eliminative,dennett1993consciousness} may hold that advanced neuroscientific theories may provide means to collate the experiences of awe eventually. However, as we will see below, collatability is a prerequisite of any theory that addresses specific aspects of conscious experience. If an aspect of experience is non-collatable, no theory can be empirically inferred or tested that addresses this aspect of experience.
\end{ExampleS} 
 
\begin{ExampleS}\em \label{ExColour} 
%

Similarly, there is at present no possibility to meaningfully ask the question of whether colour experiences of two experiencing subjects are the same or different. 
This simple but important fact is pointed at by the plain question of how two experiencing subjects might come to conclude that the experience of colour which they have if they look at, e.g., the clear sky is the same. They may ensure that they use the same reference (`blue') for the experience, that they see the same wavelength and they might even be able to conclude that similar neuronal assemblies are active in both of their brains while having the experience in question. However, none of this is a priori related to the color aspects of their experiences (`what it is like to see blue').

Put differently, there is no reasonable way to assign truth values to statements of the form `my colour experience $\e_1$ is equal to your colour experience $\e_2$', equality is not a well-defined concept when referencing to experiences of two different experiencing subjects. 
Thus colour experiences are non-collatable aspects of experience in the sense of Phenomenological Axiom~\ref{PFcollatable}.

This non-collatability has consequences for any scientific account of colour experience. E.g., any hypothesis that a particular neural activity occurs whenever a subject is experiencing a colour `green' is not well-defined, simply because there is no intersubjectively meaningful reference to `green'; the colour experience one subject is having when when presented a 510nm light source may be very different from the colour experience another subject is having when presented the same light source.
In other words, any intersubjective reference to colour experiences carries a certain {\em ambiguity}, which has to be taken into account when constructing models or designing experiments related to colour experience.
\end{ExampleS}

The main point of this paper, argued for in detail below, is that non-collatable aspects of experience cannot be addressed by the usual scientific methodology. 
Since the term `qualia' is generally used to denote what is considered as essential in a particular analysis of experience, we introduce the following abbreviation.

\begin{Def}\em \label{DefQualia}
We define the term {\em qualia} to refer to all non-collatable aspects of experience of an experiencing subject within the class $\mathcal C$.%
\end{Def}

\begin{ExampleS}\label{ExColourQualia}\em According to Example~\ref{ExColour}, colour experiences satisfy the condition of Definition~\ref{DefQualia}. Thus colour experiences are qualia.\footnote{We generally abbreviate `colour aspects of experience' by `colour experience'.}
\end{ExampleS}

\begin{ExampleS}\em \label{ExWhatIsItLike}
Example~\ref{ExColourQualia} is a special case of the aspects of experience referenced by Thomas Nagel in~\cite{Nagel.1974} when introducing his famous notion of `What is it like to be ...\,?':
\begin{quote}\em 
``[F]undamentally an organism has conscious mental states if and only if there is something that it is like to {\em be} that organism  -- something it is like for that organism. We may call this the subjective character of experience.''~(p. 436)
\end{quote}
Nagel also uses the term {\it ``how it is for the subject himself''}~(p. 440) to point to these aspects of experience.
Though~\cite{Nagel.1974} does not make the distinction of Phenomenological Axiom~\ref{PFcollatable} central to his line of reasoning, one can find hints toward this distinction in~\cite{Nagel.1974}: E.g., he claims that {\it``we do not \ch{possess} the vocabulary to describe [what it is like to be us] adequately''}~(p. 440), there are {\it ``facts that do not consist in the truth of propositions expressible in a human language.''}~(p. 441)
\end{ExampleS}

\subsection{Formal Representation of Experience}\label{SecFormal}
In order to define a formal representation of experience, we make use of two further basic phenomenological \ch{axioms}.
These are very general in nature and it is plausible that they hold independently of the particular choice of class~$\mathcal C$. However,
due to the restricted possibility of phenomenological analysis mentioned above, we generally assume $\mathcal C$ to comprise adult humans.
The first phenomenological \ch{axiom} expresses the observation that some qualia are \emph{experienced} as identical, whereas others are not, or in other words, that one sometimes experiences a non-collatable aspect as identical to a non-collatable aspect one has experienced at another time.
\begin{PA}\em \label{PFRecognize}
Qualia can be {\em recognised} to a certain \ch{extent}: Experiencing subjects can identify qualia which they have previously experienced.
\end{PA}

\begin{ExampleS}\em 
\ch{Phenomenological Axiom~\ref{PFRecognize} states that experiencing subjects may perceive some aspects they experience at different times to be identical. For example, it could be the case that someone finds the taste aspect experienced when trying artificial strawberry flavour to be identical to the taste aspect experienced when eating an actual strawberry. This recognition of previously experienced aspects is simply a ``subjective impression'' of identity, so to speak.}
\end{ExampleS}

Phenomenal Fact~\ref{PFRecognize} is important because it is the basis of the ability of an experiencing subject to introduce {\em labels} for his/her qualia, i.e. a name or reference for non-collatable aspects of his/her experience. Recognisability is presupposed in the notion of collatability, so that 
labels of collatable aspects of experience can be introduced by definition.

In what follows, we assume that labels are chosen such that different aspects of experience are associated with different labels and, using Phenomenological \ch{Axiom}~\ref{PFRecognize}, that the same label is used to denote various occurrences of the same aspect.%
\footnote{\label{Assume}Note that throughout this section, assumptions are in fact conventions. E.g., this assumption can be satisfied by asking experiencing subjects (in an experiment, say) to choose labels as described. The assumptions can be made `without loss of generality', so to speak.}
Furthermore, we assume that all experiencing subjects use the same set of labels, which we denote by $\E$. For our purposes, $\E$ can be any set, which labels the set consists of does not matter in what follows.\smallskip

The second phenomenological \ch{axiom} expresses the observation that something can be said about how non-collatable aspects occur in, or constitute, experience. In~\cite{Nagel.1974}, it corresponds to the observation that ``structural features of perception might be more accessible to objective description, even though something would be left out''~\cite[p.\,449]{Nagel.1974}.
In~\cite{Chalmers.1996}, it corresponds to the observation that 
``even if experiences are in some sense `ineffable,' relations between experiences are not; we have no trouble discussing these relations, whether they be relations of similarity and difference, geometric relations, relations of intensity, and so on. \ch{As Schlick~\cite{schlick1969form} pointed out, the \emph{form} of experience seems to be straightforwardly communicable, even if the \emph{content} (intrinsic quality) is not''~\cite[p.\,224]{Chalmers.1996}.}

\begin{PA}\em \label{PFRel} Qualia have {\em relations} that can be {\em collated} within the class $\mathcal C$.
\end{PA}

By Definition~\ref{DefPA}, this is a claim about experiences of all experiencing subjects in the class $\mathcal C$. 
In simple terms, it expresses the fact that something can be said about non-collatable aspects of experience, something about how they appear in experience.
The collatability of the relations implies that we may {\em represent the relations on the set of labels}~$\E$ and assume (i.e. ask, cf. Footnote~\ref{Assume}) labels to be chosen in such a way that the experienced relations between qualia are reflected in the relations represented on the labels.
We assume that Phenomenological Axiom~\ref{PFRel} also holds for collatable aspects of experience, so that they have relations, too, that can be represented on the set of labels.\footnote{\ch{Phenomenological Axiom~\ref{PFRel} states that there are relations between qualia which are collatable. This expresses the observations in~\cite{Nagel.1974},~\cite{Chalmers.1996} and~\cite{schlick1969form} that structural features of perception, relations between experiences or the form of experience might be more accessible to communication or objective description. However, one might question whether this axiom is warranted, and insist that relations between experiences are not (strictly, at least) collatable. (Thanks to an anonymous referee for pointing this out.) The formalism developed here requires the collatability of relations, so that any non-collatable relation has to be ignored.}}

\begin{ExampleS}\label{ExamplePFRel}\em
For qualia of the `what it  is like to be' type (introduced in Example~\ref{ExWhatIsItLike}) these relations include
\begin{itemize}[label=-,topsep=0pt]
	\itemD {\em Similarity}: Two qualia can be more or less similar.
	\itemD {\em Intensity}: A quale can occur in more or less intense versions.
\end{itemize}
among others.\footnote{Similarity and intensity are simple examples of collatable relations between qualia. There may be many more collatable relations  which express facts about how qualia appear in experience, some of which may only relate qualia of a particular type to each other. Further examples arguably include:
{\em Composition}: Some qualia are experienced as a composition of two (or more) different qualia. I.e., the composed quale is but a combination (or simultaneous experience) of the composing qualia. {\em Inclusion}: Some qualia may be experienced as containing one (or more) other qualia. Here, the contained quale is but an aspect of the containing quale. 
Also, the distinction between various {\em types} (visual, auditory, tactile, etc.) of non-collatable aspects of experience is a relation in the sense of Phenomenological \ch{Axiom}~\ref{PFRel}.}
\end{ExampleS}

\begin{ExampleS}\label{ExColourRel}\em Experiencing subjects typically experience some pairs of colours as similar to each other, whereas they experience others as not similar. E.g., small changes in hue usually result in colours which are perceived as similar, whereas large changes in hue result in colours which are not experienced as similar.

What is crucial for our purposes is that one may (and in practise often does) represent the experience of similarity of colours on the set of colour labels. Correspondingly, 
one may (and in practise often does) ask experiencing subjects to choose labels for their colour experiences in such a way that colours which are experienced as similar are similar according to the representation on colour space.\footnote{Note that this example is complicated by the fact that we {\em calibrate} colour experiences in practise: We apply or learn rules on how to pick colour labels related to external events such as wave-length impinging on the eye. This will be discussed in detail in Example~\ref{ExColour2} below. What is crucial is that a priori, individual labels so chosen do not correlate with colour experience: Two experiencing subjects may have a completely different colour experience despite using the same label `blue'.}
We will study this in detail in Example~\ref{ExColour2} below. 
\end{ExampleS}

Phenomenological \ch{Axiom}~\ref{PFRecognize} provides the possibility to introduce labels for non-collatable and collatable aspects of experience.
What Phenomenological \ch{Axiom}~\ref{PFRel} adds to this is the possibility to represent relations between aspects of experience on the set of labels. 
Since any representation of a relation on a set is mathematical in nature, so are these representations. They give either
relations on $\E$ in the mathematical sense of the word (i.e. a subset $R$ of $\E \times \E$) or some more involved mathematical structure, which turns $E$ into a mathematical \emph{space}.\footnote{\ch{Here, the term `mathematical space' is used to refer to a set which carries additional mathematical structure. Examples are metric spaces, topological spaces, vector spaces, differentiable manifolds, principal bundles, measurable spaces and Hilbert spaces.}}

Thus, together, these two phenomenological \ch{axioms} ground a representation of experience in terms of mathematical structure. 
We refer to the set of labels $\E$ together with its mathematical structure that represents relations between qualia as
\beq\label{LabelSpaceL}
\textrm{\em Experience space } \E ,
\eeq
though it is important to keep in mind that this space does not describe experience per se, but only labels and the structural relations between aspects of experiences they represent. This space $E$ is the \emph{mathematical representation of experience} mentioned above. Every element $\e \in \E$ refers either to a collatable aspect of experience or to a quale. Several detailed examples are given in Section~\ref{SecExamplesBasic} below.
\medskip

In order for the mathematics to come out right in what follows, we have to introduce an important mathematical convention with respect to collatable aspects of experience. By Definition~\ref{Noncollatable}, an aspect of experience is collatable if its identity over all experiencing subjects in the class~$\mathcal C$ can be established. This implies, in particular, that this aspect of experience can be referenced: In virtue of its collatability, it can be assigned a unique label used by all experiencing subjects in~$\mathcal C$. Our convention, in what follows, is that this is represented in the mathematical structure of $E$. 

\begin{Convention}\label{ConvColl}\em We assume that for every collatable aspect of experience, the mathematical structure of $E$ contains a unary collatable relation $\chi$ which allows one to select this aspect of experience uniquely.\footnote{A unary relation on $E$ \ch{is simply} a subset of $E$. }
\end{Convention}

\noindent \ch{In practise, this means that for any $e \in E$ which is collatable, there is a subset $\chi_e \subset E$ which contains only $e$. This convention ensures that changes of labels, discussed next, can be represented conveniently using the automorphism group. It ensures that all the collatable information is represented in the relations between aspects, so that all aspects can be treated alike in the technical definitions that follow.}

\ch{In summary, so far we have constructed a space $E$ whose elements denote aspects of experience (both collatable and non-collatable ones), e.g. phenomenal properties or elements of experience. Furthermore, this space carries relations or more advanced mathematical structures that expresses the structural features of experience, as well as information about which aspects of experience are collatable (in virtue of Convention~\ref{ConvColl}). This allows us to give a concise account of references to aspects of experience that takes non-collatability into account, as we explain next.}

\subsection{References to Qualia}\label{SecQualiaRef}

In virtue of non-collatability, any reference to qualia is ambiguous. In this section, we explain in detail why and in doing so, develop formal tools that allow us to quantify this ambiguity precisely.

\ch{We proceed in two steps. First, we discuss the case where an experiencing subject uses labels to report on his/her experience without taking into account any of the collatable relations. This is a preparatory step whose purpose is to explain the following constructions in detail. Since it ignores the relations on $E$, i.e. structural features of experience, it is artificial and will give a pathological result. Subsequently, in the second step, we discuss the appropriate case which takes the mathematical structure of $E$ into account.}

\noindent {\em Preparation: References that ignore relations.} Let us assume that an experiencing subject uses labels to report on his/her experience without taking into account any of the collatable relations. In this case the experiencing subject is free to choose any label to denote any aspect of experience, the only requirements being that different labels are used for different aspects and that the same label is being used for a recurrent aspect.
We call a choice of labels of an experiencing subject to denote his/her experienced aspects a {\em labelling} and use the term {\em relabelling} to denote a change of labelling.
In the present case, a relabelling is simply a map 
\beq\label{RelabelingTrans}
s:  \E \rightarrow \E, \quad \e \mapsto s(\e) \: ,
\eeq
which determines which label $s(\e)$ replaces the previous label $\e$.
Since different aspects are required to carry different labels, this map is injective. Since it furthermore has domain and codomain~$\E$, it is bijective.
Since any composition of two relabellings of the form~\eqref{RelabelingTrans} yields another relabelling, and since due to the bijectivity, each map~\eqref{RelabelingTrans} is invertible,
all possible relabellings form a group: The group of all bijective maps from~$\E$ to itself. This group is called the {\em symmetric group} of the set~$\E$.

The crucial insight here is that the group of relabellings allows \ch{us} to quantify the ambiguity of any statement that refers to aspects of experience. Consider e.g. the case where a statement only involves one label $\e_1 \in \E$. Since we are disregarding collatable relations at this point, this statement could just as well have been formulated with any other label $\e_2 \in \E$, simply because an experiencing subject may choose any label whatsoever to denote any quale. Mathematically, this is reflected by the fact that there is at least one relabelling~$s$ such that $s(\e_1) = \e_2$.  The same reasoning can be applied to sequences $(\e_1, ... \, ,\e_n)$ of labels, 
e.g. obtained by verbal reports at subsequent times. The ambiguity of a sequence $(\e_1, ... \, ,\e_n)$ of labels is the set of all sequences $(\e_1', ... \, ,\e_n')$ which can be obtained from the former by a relabelling $s$, i.e. the set of all sequences $(\e_1', ... \, ,\e_n')$ for which there exists a relabelling $s$ such that $(\e_1', ... \, ,\e_n') = (s(\e_1), ... \, ,s(\e_n))$.

These statements are in fact statements about equivalence classes. To see this, define two labels $\e_1$ and $\e_2$ to be equivalent, $\e_1 \stackrel{.}{\sim} \e_2$, if and only if there exists a relabelling $s$ such that $s(\e_1) = \e_2$. The ambiguity of a label $\e \in \E$ is given precisely by the equivalence class of this label,
\begin{align}\label{EqClassNC}
[\e] := \big\{ \e' \, | \, \e' \stackrel{.}{\sim} \e \big \} = \big\{ \e' \, | \, \exists \,s : \e' = s(\e) \big \} \:,
\end{align}
because this class contains all labels which an experiencing subject could have chosen. The same is true for sequences: 
If we define two sequences to be equivalent, $ \, (\e_1', ... \, ,\e_n')  \stackrel{.}{\sim} (\e_1, ... \, ,\e_n)$, if and only if there exists a relabelling~$s$ such that
$(\e_1', ... \, ,\e_n')  = (s(\e_1), ... \, , s(\e_n) )$, the ambiguity of a sequence of labels is given precisely by the equivalence class of this sequence,
\begin{align}\begin{split}\label{EqClassNCS}
[(\e_1, ... \, ,\e_n)] :=& \big\{ (\e_1', ... \, ,\e_n')  \, | \, (\e_1', ... \, ,\e_n')  \stackrel{.}{\sim} (\e_1, ... \, ,\e_n) \big \} \: ,
\end{split}\end{align}
because this class contains precisely all those descriptions of the sequence which an
experiencing subject may give.
Another way to put this is that the equivalence classes~\eqref{EqClassNC} and~\eqref{EqClassNCS} are what is empirically well-defined, not the labels themselves,
these only have meaning for the experiencing subject him/herself once he/she has chosen a particular labelling.

\ch{This concludes the description of the case that ignores structural features of experience. Its artificial nature is reflected in the fact that the symmetric group allows one to map any choice $(\e_1, ... \, , \e_n)$ of labels to {\em any} other choice $(\e_1', ... \, , \e_n')$, provided that every label occurs at most once in each choice. Thus there are very few equivalence classes (only one if $n=1$). We now proceed to the discussion of the appropriate case.}

\noindent {\em Taking Relations into Account.}
Next, we take into account the collatable relations between aspects of experience as established in Phenomenological Axiom~\ref{PFRel}.
To do so, we work with the {\em experience space}~$\E$ introduced above: I.e., we assume that the relations between qualia have been
represented on the set of labels%
\footnote{For all practical purposes, one can obtain such a representation by simply asking one experiencing subject to pick a labelling and to report, in terms of this labelling, on his/her experienced relations. Other experiencing subjects are then required to choose labels according to this representation. For details, see Example~\ref{ExColour2} below. For explicit examples on how such a representation might look, cf. Examples~\ref{ExPreTop} to~\ref{ExHS} below.}
and ask experiencing subjects to pick labels for the qualia they experience in accordance with this representation. As above, we refer to any such choice as {\em labelling}.

This constraint on how labels can be chosen implies that the freedom of every experiencing subject to choose labels is smaller than in the case above: Functions~\eqref{RelabelingTrans}
only constitute relabellings if they preserve the collatable relations represented on the set~$\E$, i.e. if they preserve the structure of the space~$\E$.
A bijective function from a space to itself which preserves the structure of this space is called an {\em automorphism} of the space. As above, automorphisms form a group. Thus in the case where we take into account the collatable relations, {\em relabellings} are elements of the 
\beq\label{AutGroup}
\textrm{\em Automorphism group}\,\Aut(\E).
\eeq
We summarize this by saying that the automorphism group~$\Aut(\E)$ describes the {\em freedom of relabelling} of every experiencing subject.

It is here that Convention~\ref{ConvColl} is important. Since for every collatable aspect of experience there is a unary relation which the automorphism group needs to leave invariant, this convention ensures that automorphisms do not change labels of collatable aspects of experience. As a result, the automorphism group
allows us to quantify precisely the ambiguity inherent any reference to qualia.

In order to identify the ambiguity of any statement that uses a sequence $(\e_1, ... \, ,\e_n)$ of labels, we can argue exactly as in the simplified description above, replacing transformations~\eqref{RelabelingTrans} by automorphisms. The result is that the ambiguity is given precisely by the equivalence class
\begin{align}\begin{split}\label{EqClassC}
[(\e_1, ... \, ,\e_n)] :=& \big\{ (\e_1', ... \, ,\e_n')  \, | \, (\e_1', ... \, ,\e_n') \sim (\e_1, ... \, ,\e_n) \big \} \: ,
\end{split}\end{align}
where $\sim$ denotes the equivalence relation defined as 
\begin{align}\label{AutEquiv}\begin{split}
 \, (\e_1', ... \, ,\e_n') \sim (\e_1, ... \, ,\e_n) \quad &\textrm{ if and only if there is an }s \in \Aut(\E) \\
 &\textrm{ such that } \e_i' = s(\e_i) \textrm{ for all } i = 1, ... \, , n \: .
\end{split}\end{align}
This class contains precisely all descriptions of the sequence of aspects of experiences which an experiencing subject might give:  A description in every possible labelling. To obtain the ambiguity of individual labels, we simply set $n=1$.

In summary, what this shows is that the empirically well-defined references to experience are given by elements of the quotient space
\beq\label{Intersub}
	\bigslant{\E^{\times n}\!}{\sim} \:,
\eeq
where $\E$ is the space~\eqref{LabelSpaceL} whose structure represents collatable relations between aspects of experience, where $\sim$ denotes the equivalence relation~\eqref{AutEquiv} and where $n \in \N$ is the length of a sequence. 

\begin{Remark}\label{ExternalReference}\em
In practise, we typically establish labels by reference to particular ``external'' events, such as particular wavelengths emerging from a light source in the case of colour experiences. Socially established labels of this sort are of course very useful in various circumstances, precisely because they correlate with external events. However, a priori there is no reason to assume that qualia of different experiencing subjects which are denoted by the same label are the same, even if the labels correlate with the same external event.\footnote{One may even take this to be unlikely, given the difference of brain physiology and neuronal structure across individuals.} 
In fact, an assumption of this kind has no empirical meaning because the definition of qualia implies that neither the identity of qualia of different experiencing subjects with an external event, nor the equality of qualia of different experiencing subjects can be empirically tested. Statements of this sort can only be meaningful if formulated based on a scientific methodology which is compatible with the non-collatability of the aspects of experience under consideration.~\QEDrem
\end{Remark}

\subsection{A Phenomenological Grounding of the Scientific Study of Consciousness}\label{SecPhenGr}

In the previous sections, we have fixed basic terminology, such as what we take the term experience to denote and how qualia are defined in these terms. We have furthermore used phenomenological \ch{axioms} to warrant introduction of labels, which has in turn allowed us to ground a mathematical representation of experience. Finally, we have analysed the implications of non-collatability (and hence of {ineffability}, {privateness} and {inaccessibility}) in terms of this formal representation.
Together, this gives rise to a grounding of the scientific study of consciousness, i.e. allows us to specify what is to be studied and how. 

First, concerning the task of the scientific study of consciousness, what is to be studied is simply experience as defined in Definition~\ref{DefExperience} and its relation to the physical domain. By Phenomenological \ch{Axiom}~\ref{PFcollatable}, this includes collatable aspects of experience as well as qualia.
What is required to do so is a combination of the usual scientific methodology with some novel tools (developed in the remainder of this article).
How these methodologies are combined is described by the formal representation of conscious experience in the experience space $\E$.

The usual scientific methodology, e.g. the one in use today in the neuroscience of consciousness, can be applied to all intersubjectively well-defined references to experience, i.e. to the equivalence classes~\eqref{EqClassC}.
When taken together, they constitute the quotient space~\eqref{Intersub}, which provides a comprehensive description of all
intersubjectively meaningful aspects of experience. This quotient space contains, in particular, all references to collatable aspects of experience.

However, the usual scientific methodology cannot be used to investigate individual elements of equivalence classes~\eqref{EqClassC}, if a class has more than one element, because the experiences labelled by these elements cannot be referenced intersubjectively in a meaningful way. These elements in fact generate an explanatory gap (Section~\ref{ExplGap}).
The study of these aspects of experience is, nevertheless, part of the task of the scientific study of consciousness. There is a fact as to which member of an equivalence class is experienced, and this fact cannot a priori be excluded from constituting a scientific explanandum.

The main achievement of this article in the following sections is to show that formal tools can be defined that allow us to go beyond a scientific analysis of the quotient space~\eqref{Intersub}. Referring to these results, we can specify the grounding that arises from the previous definitions as follows. Due to the importance of phenomenological \ch{axioms} in grounding the formal structure, we refer to this grounding as phenomenological grounding.

\begin{Def}\label{DefTask}\em 
{\em What is to be studied} by the scientific study of consciousness according to the {\em phenomenological grounding} is experience as defined in Definition~\ref{DefExperience} and its relation to the physical domain.
This includes the study of intersubjectively well-defined aspects of experience using the quotient space~\eqref{Intersub} and \emph{standard scientific methodology}, as well as the study of qualia proper, represented formally in the experience space~\eqref{LabelSpaceL} using the \emph{formal-mathematical methodology} derived in Section~\ref{PhenGrMethod}.
\end{Def}

\subsection{Examples}\label{SecExamplesBasic}
We close this section with several examples. First, in Example~\ref{ExColour2}, we continue the discussion of colour experience and show that colour spaces, which are largely in use in commercial applications, constitute the experience spaces for colour qualia as defined above. In Example~\ref{ExPreTop} to~\ref{ExHS} we consider various possible mathematical structures of the experience space~$\E$, some of which have been proposed in the literature.

\begin{ExampleS}\em \label{ExColour2}
To illustrate the meaning of the experience space~$\E$ and the group~$\Aut(\E)$, as well as Remark~\ref{ExternalReference}, we consider again colour experiences.
As we have explained in Example~\ref{ExColour}, these satisfy the defining property of qualia.

We will generally denote the quale `what it is like to see light of wavelength $\lambda$' as `experience of $\lambda$'. For the purpose of this example, we will disregard of the fact that colour experience is highly sensitive to the geometry of the lighting of a scene and to the expected material properties of an object's surface. We will use the symbol $\bar \lambda$ to denote a mixture of light of varying wavelength.
\newcommand{\Lcl}{\E_{\textrm{cl}}}

We start by fixing a particular human oberver, the ``standard observer''~\cite{Kuehni.2010}, and choose a set $\Lcl$ that is in one-to-one correspondence to all colours which this human can experience.
As usual in colour science, we assume that there is a large class~$\mathcal C$ of humans which have the same set of possible colour experiences as the standard observer. This assumption implies that every human in the class~$\mathcal C$ can specify a one-to-one mapping between the set~$\Lcl$ and his/her colour experiences.  
The fact that color experiences are qualia as defined here is reflected in the fact that there is no \emph{unique} one-to-one mapping. The set~$\Lcl$ is thus a set of labels of colour qualia as introduced after Phenomenological \ch{Axiom}~\ref{PFRecognize}. It is also the basis of the definition of colour spaces (cf. below).

The set~$\Lcl$ can be calibrated: Since colour experiences arguably arise as a response to mixtures $\bar \lambda$ of light impending on the retina, we may identify every element~$\e \in \Lcl$ with a particular mixture $\bar \lambda$.
The set of mixtures visible to the human eye can, in turn, be represented%
\footnote{This is an experimental fact which is due to biological details of the cone cells in the human eye.
Since various mixtures~$\bar \lambda$ evoke the same colour experience, some conventions have to be made in order to fix the subset~$S$ uniquely (e.g. a choice of reference wavelengths). Also, due to the particular responsivity curves of the cone cells, no finite set of wave-lengths can be combined to achieve all colours that a human can experience. However, suitable experimental procedures exist so that all visible mixtures can be represented in~$\R^3$ nevertheless~\cite{Kuehni.2010}.}
as a subset $S\subset\R^3$, roughly speaking by taking the three components of a vector $v \in S$ to represent the relative intensities of three reference wavelengths. Putting these two steps together, we may in fact choose the set~$\Lcl$ to be the subset $S \subset \R^3$. 
In this case, every label $\e \in \Lcl $ is a $3$-tuple of real numbers which specifies which mixture~$\bar \lambda$ of light has to be presented to a particular human to evoke the quale that he/she has {\em denoted} by that very label $\e$.

This calibration may lead one to think that there is a unique way of referring to colour qualia. However, this is not the case. To see this, assume that we fix some label/vector $\e \in S = \Lcl$ as well as two experiencing subjects~$A$ and~$B$. Let us denote the mixture of light that corresponds to this vector as $\bar \lambda_v$. When we present this mixture~$\bar \lambda_v$ to the two experiencing subjects, subject~$A$ has the colour experience he/she has labelled as~$\e$, and so does subject~$B$. However, this has nothing to say on whether the colour experiences are the same or not: E.g., subject~$B$ might have the colour experience subject~$A$ is having upon presentation of a completely different mixture~$\bar \lambda_w \neq \bar \lambda_v$.

This illustrates the fundamental difficulty related to qualia as defined in Definition~\ref{DefQualia}: If we would ``know'' (e.g. as the result of some scientific investigation) that the presentation of the same colour stimuli~$\bar \lambda_v$ to various subjects results in them having the same colour experience, we could meaningfully talk, or refer to, colour experiences of different subjects in terms of stimuli.
More generally, if statements of the type 
\beq\label{ExColourStatement}
	\textrm{``subject $A$ will have colour experience $X_1$ once presented input $\bar \lambda$''}
\eeq
would be known, these statements would allow us to directly refer to $A$'s colour experiences, putting us into the position to do science as usual. However, the fundamental difficulty of the subject is that statements like~\eqref{ExColourStatement} do not carry any intersubjective meaning at all: Due to the impossibility of collating colour experiences, statement~\eqref{ExColourStatement} 
cannot be distinguished (by anyone but subject $A$) from the statement
\[
	\textrm{``subject $A$ will have colour experience $X_2$ once presented input $\bar \lambda$''},
\]
where $X_2$ is any colour experience of $A$ with the same unary collatable relations (such as intensity). This problem exists independently of whether we consider the statement~\eqref{ExColourStatement} to be a hypothesis or to be the result of some purported scientific investigation. Statements of this type do not have unambiguous intersubjective meaning.

As explained above, what has intersubjective meaning are the equivalence classes~\eqref{EqClassC}. They express facts about colour experience which are invariant with respect to the labelling that an experiencing subject chooses. We now illustrate this in detail for colour qualia.

First, we need to find the collatable relations between colour experiences
referred to in Phenomenological \ch{Axiom}~\ref{PFRel}. Luckily, this has been on the agenda of colour science for decades, so that we may simply turn to its results. Put in simple terms, there seem to be three types of collatable relations~\cite{Kuehni.2010}:
Continuity of change of colours (whether some time-continuous sequence of colour experiences is perceived as continuous or not), behaviour under mixtures of colours (whether a mixture of two colour experiences is perceived as equal to another colour experience or not) and (less well known) a notion of distance of colours (whether two colour experiences are perceived as more different to each other than another pair of colour experiences).

Next, we need to translate these collatable relations into mathematical structures on the set~$\Lcl$. This yields the experience space~\eqref{LabelSpaceL} of colour qualia. Again, colour scientists have done the work for us: They have defined colour spaces in order to formalize these collatable relations~\cite{Kuehni.2010}. A colour space is a closed subset~$S$ of~$\R^3$
which is in a one-to-one correspondence with all colours humans may experience, chosen such that continuity is represented by the induced topology of~$\R^3$ (a path of colours experiences is continuous if the labels form a continuous path in the colour space), mixture is represented by straight lines (equal mixing of two colour experiences~$\e_1$ and~$\e_2$  yields the colour experience that carries the label that is at the center of the straight line that connects~$\e_1$ and~$\e_2$), and finally experienced distance of colour qualia is represented by a metric on~$S$. \ch{\footnote{\ch{We take it that straight lines describe mixtures of colour experiences, which have to be distinguished from the experience of mixtures of colours. Thanks to an anonymous referee for pointing this out.}}}
Thus a colour space is a experience space~\eqref{LabelSpaceL} for colour qualia.

There are many subsets of $\R^3$ which satisfy these requirements: For any choice of subset~$S$, there is a large class of transformation of $\R^3$ which, together with a corresponding transformation of the metric, yield another subset $S'$ of $\R^3$ which equally represents colour experiences as well as their collatable relation. Colour science uses the calibration described above to fix specific choices of subsets $S$,
so that the coordinates of the elements of $S$ can be translated into mixtures of wavelengths $\bar \lambda$. 
However, as explained above, for the study of colour {\em experience}, calibrations do not have any relevance a priori, so that no particular choice of subset can be singled out. 

In order to specify the group of relabellings for this example, we note that in more abstract terms, a colour space is\footnote{\label{ExMathColourSpace}Cf.~\cite{Kuehni.2010}. However, note that a more axiomatic treatment may result in different mathematical spaces~\cite{Resnikoff.1974,Provenzi.2017}. Furthermore, the assumption of smoothness may not be justified and one might have to consider manifolds with corners.}
 a smooth $3$-dimensional {\em Riemannian manifold}: Its topology represents the continuous changes of colour experience and its metric $g$ specifies both the geodesics (generalized ``straight lines''), which describe the mixture of colour experiences, as well as a distance function which describes the experience of distance between colour qualia. The various choices of subsets $S$ of $\R^3$ correspond precisely to choices of coordinates of this manifold. We summarize this as
\beq\label{ExRiem}
	\E = (\Lcl, g) \:.
\eeq
This is the actual form of the experience space~\eqref{LabelSpaceL} of colour qualia. Its elements label the set of colour experiences and its structure represents the collatable relations between them. 
An experiencing subject can specify his/her colour experiences by specifying points (in the case of individual colour experiences) or curves (in the case of time-continuous colour experiences) on this manifold. The freedom of choosing labels is described by the automorphism group of~$\E$. 
In the case~\eqref{ExRiem} of a Riemannian manifold, this is the group of isometries, i.e. diffeomorphisms which leave the metric invariant:
\[
	\Aut(\E) = \Iso(\E) \: .
\]
Thus the ambiguity of any statement in terms of colour labels $(\e_1, ... \, ,\e_n)$ is given by the equivalence class
\[
[(\e_1, ... \, ,\e_n)]
\]
which is defined as in~\eqref{EqClassC} with two sequences being equivalent if there is an isometry $s \in \Iso(\E)$ which transforms every element of the first sequence into the corresponding element of the second sequence.\footnote{Since the ordering of distances between pairs of colours, rather than the numerical value of the distance itself, is collatable, one could make the point that the relabelling freedom is given by the group of diffeomorphisms which leave the metric invariant up to a conformal factor. Since the present example is, mainly, of a pedagogical interest, we do not explore this further at this point. Cf. also Footnote~\ref{ExMathColourSpace}.}
 The actual form of the equivalence classes depends on the metric $g$, which can be determined experimentally. The current version of the distance function internationally in use \ch{is} reviewed e.g. in~\cite{SharmaWuDalal.2004}, a discussion of which however goes beyond the scope of this example.\footnote{We note that it is possible that some sequences $(\e_1, ... \, ,\e_n)$ are not ambiguous, i.e. that $[(\e_1, ... \, ,\e_n)] = \{ (\e_1, ... \, ,\e_n) \}$.
This means that there is one unique sequence of colour experiences which has the properties represented by the sequence $(\e_1, ... \, ,\e_n)$ of labels, or put differently, that there is only one possible choice of labels for this sequence that takes into account the collatable relations as described. Sequences of this kind may be used to remove the ambiguity of the labels they contain and make these aspects of experience accessible to a proper scientific analysis.
}

Putting everything together, we conclude that any statement, scientific or otherwise, that addresses colour experiences sensu stricto -- i.e. which addresses what it is like to experience colours -- only makes sense if it is invariant with respect to~$\Iso(\E)$ transformations.
This is a consequence of the fact that qualia are non-collatable and of the corresponding freedom of every
experiencing subject to choose names for the qualia he/she experiences.

The difference between labels of colour experiences and colour experiences (colour experiences de dicto and colour experiences de re, so to speak), can be crucial for scientific investigations. For example, if a study compares the calibrated label $\e$ that a subject reports with neural activity,
it does {\em not}  investigate the relation between neural activity and colour experience but rather the relation between neural activity and presentation of wavelengths $\bar \lambda$ to the retina. These two objects of investigation refer to completely different scientific agendas. 
\end{ExampleS}

\begin{ExampleS}\em {\bf Pretopological structure on $\E$.}\label{ExPreTop}
In the previous example, we have relied on results from colour science to provide the mathematical structure of the experience space $\E$ that represents the colour aspects of experience. The goal of this example is to illustrate in more detail how the mathematical structure of $\E$ can be defined directly in terms of relations between qualia. To this end, we consider the relation of similarity of two qualia explained in Example~\ref{ExamplePFRel}, but understood in a binary way. I.e., for the purpose of this example, we make the simplifying assumption that any two non-collatable aspects of experience (of one experiencing subject) are experienced either as `similar' or as `not similar', and ignore the experience of varying degrees of similarity. While this restriction may not be warranted in practise, we take it to be justified for pedagogical purposes.
When understood in this way, the similarity relation can be used to define a pretopological structure on $\E$ as described in~\cite{Prentner.2019} (with a slightly different goal in mind), whose presentation we now follow.\footnote{The following definitions and their relation to topology  are intuitively accessible if one thinks about open balls in a metric space such as $\R^3$, where $\circ$ is defined as overlap. We remark, however, that the construction does not give rise to a topology, as claimed in~\cite{Prentner.2019}, since the third Kuratowski closure axiom (idempotence) does not follow.}

First, we define a binary relation $R_{\circ} \subset \E \times \E$ on $\E$. If two qualia with labels $\e_1$ and $\e_2$ are perceived as similar by an experiencing subject, we define the corresponding labels to be related according to $R_{\circ}$, which we denote as $\e_1 \circ \e_2$ (i.e. $\e_1 \circ \e_2 \Leftrightarrow (\e_1,\e_2) \in R_\circ$, and similarly below). Thus $R_{\circ}$ is given directly by experience. We assume that $\e \circ \e$ for all $\e \in \E$.
Second, based on the data of $R_{\circ}$, we define another relation $R_{\leq}$ on $\E$, called ``parthood relation''~\cite{Prentner.2019} as
\[ 
\e_1 \leq \e_2 \quad \textrm{ iff } \quad 
\textrm{$ \e \circ \e_1 \Rightarrow \e \circ \e_2$} \: .
\]
Thus $\e_1 \leq \e_2$ holds iff all qualia which are similar to $\e_1$ are also similar to $\e_2$.
Third, we use the parthood relation $R_{\leq}$ to define yet another relation $R_\sim$, called ``connection'', as follows:
\begin{align*}
	\e_1 \sim \e_2 \quad \textrm{ iff } \quad &\textrm{$\exists \,  \tilde \e \in \E$ such that $\tilde \e \circ \e_1$ and $\tilde \e \circ \e_2$}\\	
	&\textrm{ as well as $\e \leq \tilde \e \Rightarrow \e \circ \e_1$ or $\e \circ \e_2$} \: .
\end{align*}
Note that $\e_1 \leq \e_2$ implies $\e_1 \sim \e_2$. We extend this notation to sets $A \subset \E$ by
defining
\[
\textrm{$\e \sim A$ iff $\e \sim \tilde \e$ for at least one $\tilde \e \in A$.}
\]
This allows us to define an operator $\pcl$, which takes a subset $A \subset \E$ to another subset $\pcl(A)$ 
which contains all qualia which are connected to at least one of the qualia in $A$:
\[
	\pcl(A) := \{ \e \, | \, \e \sim A \} \: .
\]
The operator~$\pcl$ satisfies three of the four Kuratowski closure axioms~\cite[Sec.\,3.2]{Pervin.1964},
but need not satisfy $\pcl(\pcl(A))=\pcl(A)$ for all $A \subset \E$ (idempotence). Hence it constitutes a preclosure operator,
so that $(\E, \pcl)$ constitute a pretopological space~\cite{nlab:pretopological_space}.

In order to define what constitutes a relabelling in this example, we note that a function~$f$ between two pretopological spaces $(\E,\pcl)$ and $(\E', \pcl')$ is defined to be continuous if
\[
	f( \pcl (A) ) \subseteq \pcl' (f(A))
\]
for all $A \subset \E$. The automorphism group~$\Aut(\E)$ of $(\E,\pcl)$ is the set of all continuous invertible functions $f: \E \rightarrow \E$ whose inverse is also continuous, with group operation given by function composition.

Thus we see neatly how non-trivial mathematical properties of the experience space~$\E$ can be defined directly in terms of experienced relations between qualia. The similarity relation established via Phenomenological \ch{Axiom}~\ref{PFRel} may, of course, not actually be binary: There seem to be various degrees, maybe even a continuum, of similarities of qualia.~\end{ExampleS}

\begin{ExampleS}\em {\bf Partial order on~$\E$.}\label{ExComp}
Our next example goes back to~\cite{Resende.Talk.2018}. First, we
observe that next to the two relations mentioned in Phenomenological \ch{Axiom}~\ref{PFRel},
qualia may in fact have compositional relations that can be collated: An experiencing subject may
find that the ineffable aspect of an experience he/she is having at a particular time includes an ineffable aspect he/she has had at another time. In this case, we may say that the former quale {\em includes} the latter quale.
If~$\e_1$ is the label which the experiencing subject has chosen for the former quale and~$\e_2$ is the label he/she has chosen for the latter quale, we will denote this relation between the two qualia as $\e_2 \leq \e_1 $.

By convention, we may put $\e \leq \e$ for all $\e \in \E$ (reflexivity). Furthermore, it is reasonable to hold that if both $\e_1 \leq \e_2$ and $\e_2 \leq \e_1$ for two labels $\e_1, \e_2 \in \E$, these labels actually refer to the same quale, so that $\e_1 = \e_2$ (anti-symmetry). Finally, qualia seem to satisfy that $\e_1 \leq \e_2$ and $\e_2 \leq \e_3$ imply $\e_1 \leq \e_3$ (transitivity). Therefore, this actually constitutes a partial order on~$\E$ and turns $(\E, \leq)$ into a partially ordered set. The automorphism group consists of bijective functions $f: \E \rightarrow \E$ which are order-embedding, i.e. which satisfy $\e_1 \leq \e_2$ if and only if $f(\e_1) \leq f(\e_2)$ for all $\e_1, \e_2 \in \E$.
Thus one can see nicely that the automorphism group describes the freedom of relabelling: Its elements represent changes of labels which preserve the inclusion relation between qualia.
\end{ExampleS}

\begin{ExampleS}\em {\bf Involutive semigroup structure on~$\E$.}\label{ExSemiGr}
This example also goes back to~\cite{Resende.Talk.2018}. In order to state it, note that in Definition~\ref{DefExperience}, we have defined the term `experience' with respect to instants of time. This implies that qualia (being aspects of experience) are associated to a instant of time as well,%
\footnote{The term `instant of time' may refer to experiential instants of time or to instants of time as used in physics, i.e. points $t \in \R$.}
therefore excluding a sequence of two qualia arising at two consecutive instants of time to constitute another quale. However, one might drop this restriction to instants of time, and define qualia as aspects of experience in general. Following this line of thought, one could argue that for any two qualia~$\e_1, \, \e_2$, there is another quale~$\e_3$ which is the consecutive experience of the two qualia.
One might denote~$\e_3$ as
\[
	\e_3 = \e_1 \, \& \, \e_2 \:,
\]
where the `\&' represents ``and then''~\cite{Resende.Talk.2018}. If one furthermore demands associativity, which does seem to be plausible, this defines a semigroup $(\E, \&)$.

Next, one may consider an operation which reverses this temporal order of qualia. This may or may not have deep conceptual meaning: On the one hand, it may merely map any quale of the form $\e_1 \, \& \, \e_2$ to a quale of the form $\e_2 \, \& \, \e_1$, both of which have to exist due to the semigroup structure introduced above. On the other hand, it may express a deep fact about reversal of psychological time~\cite{Resende.Talk.2018}. In both cases, skipping over a few technical details, this gives rise to an involution~\cite{Resende.Talk.2018}, i.e. a map
\begin{align*}
	*: & \, \E \rightarrow \E						& \textrm{such that}\\
	&\e \mapsto \e^\ast				& (\e^\ast)^\ast = \e \, .
\end{align*}
In summary, the time composition relation of qualia may be represented on the space of labels in terms of an involutive semigroup structure.
\end{ExampleS}

%

\begin{ExampleS}\em {\bf Hilbert space structure on $\E$.}\label{ExHS}
The last example is intended to evaluate in how far the axioms of a Hilbert space can be grounded in the relations introduced in Phenomenal Fact~\ref{PFRel}. The upshot is that whereas some of the axioms can be motivated based on Phenomenological \ch{Axiom}~\ref{PFRel}, others cannot.
Nevertheless, the example may prove valuable for constructing toy models of consciousness, which is why we include it here.

In what follows, we make several assumptions about the set of all experiences which an experiencing subject might have. These assumptions are phenomenological in flavour, yet some may ultimately not be justified. 
\bei
\item[(A1)] We assume that with respect to any two qualia of one experiencing subject, the experiencing subject might have an experience which has exactly these qualia as ineffable aspects.
\eni
With respect to qualia of the `what it is like to be' type (Example~\ref{ExWhatIsItLike}), this assumption amounts to the following statement: If an experiencing subject has made an experience which included an ineffable `what is it like to be' aspect (quale) which he/she labels by $\e_1$, and another experience which included an ineffable `what is it like to be' aspect (quale) which he/she labels by $\e_2$, then it is possible that he/she will make an experience which has exactly $\e_1$ and $\e_2$ as ineffable aspects. We will use the term `simultaneous experience of $\e_1$ and $\e_2$' as an abbreviation for the statement that the experiencing subject in question has an experience which includes both aspects $\e_1$ and~$\e_2$.
To give an example, let $\e_1$ refer to what it is like to taste cheese and $\e_2$ refer to what it is like to smell wine. In this case, Assumption~(A1) amounts to granting the possibility of the experiencing subject in question simultaneously experiencing what it is like to taste cheese and what it is like to smell \ch{wine}. Whether this experience actually arises when the subject eats cheese and drinks \ch{wine} is of no concern with respect to Assumption~(A1). We take the combination of the same experience $\e$ with itself as denoting the experience of quale $\e$ but twice as intense (cf. below).
In order to motivate a group structure with respect to simultaneous experience, the following assumption is necessary:
\bei
\item[(A2)]We assume that there is a unique neutral quale which we denote by `$0$'. Furthermore, we assume that for every quale $\e$, there is \ch{a} quale $-\e$ such that an experience which includes both $\e$ and $-\e$ is not distinguishable from (and hence equal to) the experience of the neutral quale.
\eni
It seems that this assumption is utterly beyond empirical justification, since it invokes something like ``cancellation'' of `what is it like to be' aspects of experiences, so that we may only be able to ground a semigroup-structure of qualia with respect to combination (`simultaneous experience').
For the purpose of this example, we proceed nevertheless.
We denote the simultaneous experience of two qualia $\e_1$ and $\e_2$ by $\oplus$, so that the ineffable aspect of the experience which comprises both qualia labelled as $\e_1$ and as $\e_2$ established by Assumption~(A1) is labelled by $\e_1 \oplus \e_2$. Associativity and commutativity hold, so that we have:
\bei
\itemD (A1) and (A2) imply that $\oplus: \E \times \E \rightarrow \E$ is an abelian group.
\eni

Next, we model changes of intensity, as conceded in Phenomenological \ch{Axiom}~\ref{PFRel}, by a positive real number in the following sense: If $\e_2$ is the same quale as $\e_1$, but $c$ times more intense, then we denote $\e_2 = c \e_1$, where $c \in \R^+$. For $c \in \R^-$, $c \e_1$ is the opposite experience $-\e_1$ introduced in (A2), but experienced $|c|$ times as intense as $-\e_1$, where $|c|$ is the modulus of $c$. Finally, we assume that as intensity decreases, $c \rightarrow 0$, any experience goes over to the neutral quale, formally $\lim_{c \rightarrow 0} c \e = 0$ for any $\e \in \E$, where $0$ denotes the neutral quale introduced in~(A2). Making the idealized assumption
that a continuum of more and less intense versions of any experience is possible, we have:
\bei
\itemD The intensity relation of Phenomenal Fact~\ref{PFRel} may be taken to give rise to a scalar multiplication $\odot: \R \times \E \rightarrow \E$.
\eni \noindent
As usual, we suppress the symbol $\odot$ for scalar multiplication. We need to check whether the axioms of a vector space relating scalar multiplication and addition hold. Our interpretation implies that $1 \e = \e$, hence neutrality of $1 \in \R$ holds. The two axioms of distributivity read
\begin{align}
&c \, ( \e \oplus \e') = (c \e) \oplus (c \e')  &&\textrm{and } \label{AxDist1}\\
&(c+c') \, \e = c \e \oplus c' \e	&&\textrm{for all } c,c' \in \R \textrm{ and } \e,\e' \in \E \label{AxDist2}
\end{align}
Axiom~\eqref{AxDist1} says that a $c$ times more intense simultaneous experience of $\e$ and $\e'$ arises as the combination of $c$ times more intense experiences of $\e$ and $\e'$, respectively, which we take as a plausible assumption in the context of this example. Axiom~\eqref{AxDist2} sates a compatibility of addition of intensities with combinations of experience. E.g., it says that an experience $\e'$ which is the same as another experience $\e$ but twice as intense, $\e' = 2 \e$ can arise as the simultaneous experience of the combination of $\e$ with itself. We render this axiom at least somewhat plausible by defining the combination of an experience with itself to be the same experience experienced twice as intense. 
Finally, we note that the associativity axiom $(c\, c') \, \e  = c \, ( c' \e) $ is compatible with our interpretation of $\oplus$ and $\odot$. We therefore have:
\bei
\itemD $(\E, \oplus, \odot)$ satisfies the axioms of a vector space.
\eni

It remains to implement the the relation of similarity between qualia. As before, we idealize and assume that there is a non-negative real number which specifies how similar two qualia $\e_1$ and $\e_2$ are. We denote this number by $\la \e_1 , \e_2 \ra$. If $\e_1, \e_2$ are not similar at all,
we set $\la \e_1 , \e_2 \ra = 0$. If they are similar to some degree, we have $\la \e_1, \e_2 \ra > 0$, where a larger value implies more similarity.
It seems natural to impose symmetry, $\la \e, \e' \ra = \la \e', \e \ra $ for all $\e, \e' \in \E$. An inner product furthermore satisfies
\begin{align*}
&\la \e, \e \ra = 0 \, \Leftrightarrow \e = 0 && \textrm{(Definiteness)}\\
& \la \e, c \, \e' \ra = c\, \la \e, \e' \ra && \textrm{(Linearity)}\\
& \la \e, \e' \oplus \e'' \ra = \la \e, \e' \ra + \la \e , \e'' \ra && 
\end{align*}
for all $c \in \R \textrm{ and } \e, \e', \e'' \in \E$. Out of those three axioms, only the last one seems reasonable to some \ch{extent}. It says that similarity is compatible with simultaneous experience: The similarity between a quale $\e$ and the simultaneous experience of qualia $\e'$ and $\e''$ is given by the sum of the similarity of the quale $\e$ to each one of the qualia $\e'$ and $\e''$.\\ \noindent
Definiteness says that the only quale which is not similar to itself is the neutral quale. This seems rather problematic if one chooses the interpretation of $0$ introduced in~(A2). The first axiom of linearity says that the similarity between a quale $\e$ and a $c$ times more intense version of a quale $\e'$ is given by $c$ times the similarity between $\e$ and $\e'$. As mentioned before, in order to have a nice and clear example, we will accept also these assumptions for now, so that in summary we have:
\bei
\itemD $(\E, \oplus, \odot, \la . , . \ra)$ satisfies the axioms of a inner product space or Pre-Hilbert space.
\eni\noindent 
The inner product $\la . , .\ra$ introduces a norm on $\E$ as usual by $\| \e \| = \sqrt{\la \e , \e \ra}$. This norm may be interpreted as the intensity of a quale $\e$.

The inner product space $(\E, \oplus, \odot, \la . , . \ra)$ may not be complete with respect to this norm, meaning that there are Cauchy sequences in $\E$ which do not converge to an element in $\E$. In terms of qualia, this means that there are sequences of qualia whose elements become ever more similar to each other but which do not converge to any quale in the topology specified by the similarity relation. In order to exclude such cases, we consider the completion of $\E$ with respect to the norm $\| . \|$, which is unique up to isometric isomorphism. Alternatively, we may assume that there is a finite number of classes of non-similar qualia, so that completeness holds automatically. A complete inner product space is a Hilbert space. Denoting, as usual, completion by a line over the corresponding quantities, we have:
\bei
\itemD The experience space $\E$ carries the structure of a real Hilbert space $\overline{ (\E, \oplus, \odot, \la . , . \ra)}$, which we denote by $\H_\E$.
\eni
Note that this is an {\em abstract} Hilbert space: Due to the ineffability of qualia, the elements of the Hilbert space do not have an intrinsic collatable nature (as e.g. the case if one considers function spaces). 
The automorphism group~$\Aut(\E)$ is the group~$U(\H_\E)$ of unitary operators.~\end{ExampleS}

\section{Explanatory Gap}\label{ExplGap}
An ``explanatory gap''~\cite{Levine.1983} between a phenomenon%
\footnote{Here, by `phenomenon', we mean anything that occurs or manifests itself in a general sense, including both scientifically observable ``empirical phenomena'' (such as data of an experiment) as well as what is directly or indirectly perceived (experiences).}
and natural science occurs if the phenomenon has properties which render it incompatible with all notions of explanation used in natural science.
This is in particular the case if the phenomenon violates a necessary condition for the application of any of these notions of explanation.

Explanatory gaps are taken by some to indicate or entail ontological gaps (cf.~\cite[Ch.\,5, Sec.\,3.4]{Chalmers.2010}). 
Whether this is legitimate or not is a question which we will not need to address here. What matters for us is that if there is an explanatory gap, a change of methodology is necessary if the phenomenon is to be addressed by scientific means.%
\footnote{This is not to say, of course, that every phenomenon {\em can} be addressed by scientific means. There may be phenomena to which the scientific method cannot be applied. However, it seems that the only way to establish whether this is the case for a particular phenomenon is to try to develop a suitable methodology and, if successful, to apply it.}
This change may or may not be motivated by ontological considerations.

Whether there is an explanatory gap or not strongly hinges on what one takes scientific explanation to be. E.g., in~\cite{Chalmers.1996}, it is assumed that explanations in natural science can address 
``{\em only} structure and function, where the relevant structures are spatiotemporal structures, and the relevant functions are causal roles in the production of a system's behavior''~\cite[p.\,105f.]{Chalmers.2010}, which implies that phenomenal experience, defined in~\cite{Chalmers.1996} to consist of precisely those aspects of experience which do not have a  structure and function, cannot be explained by natural science. (For details, cf. Appednix~\ref{ChalmersGrounding}.) While this axiomatic derivation of an explanatory gap is undoubtedly important, the underlying notion of explanation is too narrow (Appendix~\ref{ChalmersExpl}). This calls both the explanatory gap and the grounding built on it into question.

This is different for qualia as defined in Definition~\ref{DefQualia}. Since the deductive-nomological model of explanation, the deductive-statistical model of explanation, the statistical relevance model of explanation and the causal mechanical model of explanation all tacitly presuppose that descriptions of the explanandum can be collated~\cite{Woodward.2017},
a thorough explanatory gap exists between any scientific explanation and qualia as defined here.
No scientific methodology in applied to date can be used to address non-collatable aspects of experience.

Put in simple terms, this comes about from the fact that all explanations used in natural science to date
need to assume that the phenomenon under investigation is intersubjectively accessible. Since our definition of qualia comprises those aspects of experience which are not intersubjectively accessible, they cannot be addressed by the standard methodology.
There is no possibility at present in natural science to explain why an experiencing subject experiences a particular quale over and above explanation of the collatable relations between qualia.

Thus we conclude that a change of methodology is necessary if qualia are to be addressed scientifically. 
The remainder of this article is devoted to developing this change in methodology. Our results show that mathematical
tools can be devised which allow us to address it. The resulting methodology 
generalizes Chalmers' strategy (outlined in Section~\ref{ChalmersGrounding}) and constitutes a formal framework for models of consciousness. 

\section{The Mathematical Structure of Models of Consciousness}\label{SecMathStrMoc}

Models of consciousness are hypotheses about how conscious experience and the physical domain relate.
In this section, we describe the general mathematical structure these models may take making use of the 
minimally sufficient mathematical structure of any formal scientific theory (Section~\ref{Theories}) and of the epistemic asymmetry of conscious experience (Section~\ref{PreMods}). Finally, we introduce notation that will be used further below. (Section~\ref{MocNotation}).

\subsection{Mathematical Structure of Scientific Theories}\label{Theories}

There are various different accounts in philosophy of science of what constitutes a scientific theory. Roughly, one may distinguish syntactic accounts, semantic accounts and pragmatic accounts~\cite{Winther.2016}, which differ mainly in the role they attribute to mathematical formalization. Which account of scientific theories is most adequate for \ch{the scientific study of consciousness} is yet to be seen. 

The following list of formal ingredients is general enough to include any of the above-mentioned accounts of what constitutes a scientific theory. In preparation, we remark that a family $(d_t)_{t \in \I}$ is a function $f: t \mapsto d_t$, which we will call ``trajectory''. It describes the change of dynamical variables with respect to the parameter $t$. Needless to say, the following list is not intended to be sufficient.

\begin{Def}\em\label{DefTheories} The mathematical structure of a scientific \emph{theory} $T$ comprises at least:
\bei
\itemD A set of {\em dynamical variables} $d$. (Those quantities whose variation is determined by $T$ to some extent.)\footnote{We use the word `variable' in a general sense here: A variable may represent something as simple as a natural number just as well as an operator-valued field on some manifold.}
\itemD Some {\em background structure} $b$. (Variables, or general mathematical structures, whose change is not determined by $T$. Background structure needs to be fixed in
order to determine the variation of $d$ in a particular application.)
The variation or change of the dynamical variable of a theory can be expressed with respect to some parameter $t$ which takes values in some set $\I$. Typically, the parameter is assumed continuous and interpreted as time. However, this is not necessary: The set $\I$ may or may not carry some mathematical structure (such as a topology) and it may or may not be interpretable as time. 
\itemD A set of {\em kinematically possible trajectories} $\K$. Sometimes, this includes all possible trajectories, $\K = \{ (d_t)_{t \in \I} \}$, but in many cases, trajectories need to satisfy certain mathematical requirements, such as differentiability with respect to the parameter $t$.
\itemD Some {\em laws} $\L$. (Typically equations or variational principles, but $\L$ may also include different formal ingredients (such as those provided by category theory) or even non-formal ingredients, as required by pragmatic accounts of scientific theories.)
\itemD A set of {\em dynamically possible trajectories} $\D$ which we also call {\em solutions} of~$T$. These are those kinematically possible trajectories ($\D \subset \K$) which are selected by the theory's laws in a particular application of the theory, given some choice of background structure and possibly taking into account some ``nonformal patterns in theories''~\cite[p.\,55]{Craver.2002}.
\eni
\end{Def}

In the next section, we will put these ingredients of a scientific theory into connection with the definitions introduced in Section~\ref{PhenomenologicalGrounding}. In doing so, we will have to distinguish between a general theory $T$ and those theories which have been put forward (or are anticipated) by contemporary natural science. Similar to Chalmers' use of the term `physical domain' (Appendix~\ref{ChalmersGrounding}) we will refer to the latter as {\em physical theories}. We will use the symbol $T_P$ to indicate one of these theories and denote its dynamical variables, background structure, kinematically possible trajectories and solutions by $d_P$, $b_P$, $\K_P$ and $\D_P$. 
Finally, we will assume that the physical theories are formulated in terms of a {\em state space} $P$, which is chosen such that according to the laws of $T_P$, each $p \in P$ determines a unique trajectory in $\D_P$. I.e., there is a one-to-one correspondence between solutions $(p_t)_{t \in \I} \in \D_P$ and states $p \in P$.
\smallskip

We use the term {\em model} to denote a theory which is being proposed. This includes full-fledged theories which have not received the kind of empirical support usually required in science, but also ``toy-models'', which do not aim for a comprehensive account of some class of phenomena, but rather serve to study some specific aspect of it or to test a general idea of how the phenomena could be modelled.\smallskip

Finally, for use in Section~\ref{PhenGrMethod}, we review the general definition of a symmetry group.
Note that $\K$ denotes the kinematically possible trajectories introduced above.\\[-1.5em]

\begin{Def}\em \label{DefSym}
A group $\G$ is a {\it symmetry group} of a theory $T$~\cite[p.\,43]{Giulini.2009} if and only if the following conditions are satisfied:
\bei
\item[(a)] There is an effective%
\footnote{\label{EffectiveAction}An action is effective ($\equiv$ faithful) if and only if no group element other than the identity fixes all elements of $\K$.}
action $\G \times \K \rightarrow \K$ of $\G$ on $\K$.
\item[(b)] This action leaves the the solutions $\D$ of $T$ invariant.
\eni
If $\phi$ is an action of $\G$ on $\K$ which satisfies the requirements (a) and (b), the pair $(\G, \phi)$ is a {\em symmetry} of~$T$.
\end{Def}%

\subsection{Models of Consciousness}\label{PreMods}

We now apply Definition~\ref{DefTheories} to give a general account of what a model of consciousness is. To this end, 
we make use of the epistemic asymmetry of conscious experience.

\textit{Epistemic asymmetry} is name of the most fundamental epistemological problem associated with conscious experience, namely that there ``two fundamentally different methodological approaches that enable us to gather knowledge about consciousness: we can approach it from within and from without; from the first-person perspective and from the third-person perspective. Consciousness seems to distinguish itself by the privileged access that its bearer has to it''~\cite{metzinger1995problem}. The epistemic asymmetry implies that there are two epistemically distinct notions of state, one associated with the third person perspective and one with the first person perspective.

Whereas metaphysical theories of mind may deal with only one of them, and leave the relation to the other somewhat implicit, models of consciousness may not. Being \emph{scientific} hypotheses about how experience relates to the physical domain, the relation of these two epistemically different states is exactly what models of consciousness need to address. Even if they take
the third-person state to be fundamental (as in physicalist ontologies), they \ch{need to} give a description of how the first-person state evolves in time, i.e. why conscious experience appears to be what it is. And even if they take the first-person state to be fundamental (as in idealist ontologies), they need to give a description of how the third person state evolves, i.e. why the outside world appears to be what it is. The existence of these descriptions are what marks the difference between formal ideas and scientific models.

The mathematical representation of phenomenal consciousness developed in Section~\ref{NewGrounding} is precisely what describes the first-person perspective in formal terms, with first person states being elements $\e$ of the experience space $\E$. 
A formal account of third-person states, on the other hand, is provided by natural science, which is devoted to the study of phenomena in the third person perspective in the first place. Referring to theories of natural science by and large as `physical theories', the third-person states are thus the states utilized in physical theories, e.g. states of neural networks or other descriptions of the human brain. Using the notation introduced in the last section, we denote physical theories by $T_P$ and their state space by $P$. 

\emph{In summary}, the above shows that in virtue of consciousness' epistemic asymmetry, a model of consciousness needs to prescribe a relation between states of experience and physical states, independently of which ontology it seeks to express. In formal terms, this means that it needs to prescribe a relation between the experience space $\E$ and the state space $P$ of a physical theory $T_P$. Applying the minimal formal ingredients of a scientific theory identified Definition~\ref{DefTheories}, 
this means that the dynamical variables of a model of consciousness are given by $\E \times \P$ or in fact $\E^n \times \P$ with $n \geq 1$ in case a model of consciousness can prescribe more than one experiencing subject for a given physical state (as e.g. the case with Integrated Information Theory when making use of the exclusion postulate~\cite{IIT30}).
This is summarized in the following definition.

\begin{Def}\em \label{DefPreModel} Let $T_P$ denote a physical theory. 
A {\em pre-model of consciousness}~$M$ is a theory as in Definition~\ref{DefTheories}, where:
\begin{enumerate}[label=(\roman*),topsep=0pt]
\item The dynamical variables are a Cartesian product of the physical state space $P$ of $T_P$ together with one copy of the experience space $\E$ for each experiencing subject,
\beq\label{StateSpace}
	d = 	\underbrace{\E \times \E \times ... \times \E}_{\text{experiencing subj.}} \times P \: .
\eeq
\item Kinematically possible trajectories $\K$ are a subset of families of dynamical variables,
\beq\label{defK}
	\K \subset \Big\{ \big(\e_t^1, \e_t^2, ... \, , \e_t^n,p_t\big)_{t \in \I} \, \Big\} \:,
\eeq
where $\e^i_t \in \E$, $p_t \in P$, $n$ is the number of experience spaces in~\eqref{StateSpace} and~$\I$ is some parameter space.%
\footnote{\label{Regularity}The subset $\K$ will typically be determined by demanding families $ \big(\e_t^1, \e_t^2, ... \, , \e_t^n,p_t\big)_{t \in \I}$ to satisfy some mathematical properties, such as regularity, which are necessary for the laws $\L$ of $T$ to be well-defined.
To exclude pathological cases, we assume that every label $\e \in \E$ is contained in at least one family  $(\e_t^1, \e_t^2, ... \, , \e_t^n,p_t)_{t \in \I}  \in \K$.}
\end{enumerate}
\end{Def}

We have dubbed this structure a `pre-model of consciousness' since it does not yet take into account any of the characteristic features of conscious experience, so that the above mathematical structure may as well describe any other scientific theory that addresses two variables that are epistemically distinct. An improved definition that takes into account some of the characteristic features will be given in Section~\ref{SecDefMoc} below.

Note that by referencing a physical theory $T_P$, this definition takes into account that models of consciousness are built on and extend, or allow us to derive, physical theories. An example for the former is again Integrated Information Theory (Section~\ref{ExIIT}), and example for the latter is Conscious Agent Network Theory (Section~\ref{ExCAN}). Further examples are given in Section~\ref{Examples}.

Implied by the above definition is that a model of consciousness~$M$ comes with laws~$\L$ which select from all kinematically possible trajectories $\K$ a set of solutions~$\D$. Each solution~$(\e_t^1, \e_t^2, ... \, , \e_t^n,p_t)_{t \in \I} \in \D$ consists of families $(\e_t^i)_{t \in \I}$, which describe changes of labels for every experiencing subject
$i \in \{1, ... , n \}$, and of a family $(p_t)_{t \in \I}$, which describes changes of the physical states. The solution thus realizes the mutual influence of conscious experience and physical states as described by the laws of the model~$M$.

\subsection{Notation}\label{MocNotation} We conclude this section by providing a few abbreviations that will be of use further below. We will generally use the shorthand
\beq\label{Fam}
(\bar \e_t, p_t)_t := \big(\e_t^1, \e_t^2, ... \, , \e_t^n,p_t\big)_{t \in \I}  \:,
\eeq
where $\bar \e_t = (\e_t^1, \e_t^2, ... \, , \e_t^n)$, to denote elements of $\K$. 
Furthermore, we denote by~$\D|_P$ those trajectories in the physical state space $P$ which are part of solutions $\D$ of $M$,
\begin{align}\begin{split}
\label{D-P}
	\D|_P := \big\{ (p_t)_{t \in \I} \, \big| \, (\bar \e_t , p_t)_{t \in \I} \in \D \big\} \:.
\end{split}\end{align}
This set is not necessarily equal to the set $\D_P$ of solutions of the contemporary physical theory $T_P$.
Whether $\D_P = \D|_P$ or $\D_P \neq \D|_P$ is determined by the laws~$\L$ of the model~$M$,
cf. Section~\ref{ClosureOfPhys}.
Similarly, we define
\begin{align}\begin{split}\label{K-P}
	\K|_P := \big\{ (p_t)_{t \in \I} \, \big| \, (\bar \e_t , p_t)_{t \in \I} \in \K \big\}, \\
	\K|_\E := \big\{ (\bar \e_t)_{t \in \I} \, \big| \, (\bar \e_t , p_t)_{t \in \I} \in \K \big\} \:.
\end{split}\end{align}
Since the choice of subset~$\K$ in~\eqref{defK} is a technical condition prior to the application of any law~$\L$, we may for simplicity assume that
 $\K|_P = \K_P$.

\section{Taking Characteristic Features of Conscious Experience into account}\label{PhenGrMethod}

In the previous section, we have derived the general mathematical structure of any model of consciousness. We have shown that consciousness' fundamental epistemological feature has some implications regarding the formal structure of any scientific theory that seeks to address it. However, we have not yet taken into account any of the characteristic features of conscious experience.

The goal of this section is to do so. To this end, we work with the notion of non-collatability introduced in Section~\ref{NewGrounding}. Since non-collatability is implied by {ineffability}, {privateness} and {cognitive, linguistic and communicative inaccessibility}, the mathematical structure identified here is in fact a consequence of all of these characteristic features.

Section~\ref{SecNecWellDef}, we drive formal mathematical structures of models of consciousness which are necessary to account for non-collatability. In Section~\ref{SecDefMoc}, we use the result to give an improved definition of what constitutes a model of consciousness, and show by means of an example that these structures are also sufficient to address non-collatable aspects of experience. Finally, in Section~\ref{Comparison}, we compare the improved definition of a model of consciousness with the direct description of qualia that may otherwise be used, and show that when in comes to non-collatable aspects of experience, mathematical models can achieve more than the direct description.

Together, these results show that the formal-mathematical tools developed here can in fact address the explanatory gap between qualia and natural science identified in Section~\ref{ExplGap}. Therefore, this section provides the methodology for the grounding of the scientific study of consciousness outlined in Section~\ref{SecPhenGr} when it comes to non-collatable aspects of experience.

\subsection{Non-Collatability implies Symmetry}\label{SecNecWellDef}

In Section~\ref{PhenomenologicalGrounding}, we have discussed intersubjectively meaningful references to qualia. We have found that sequences of labels $(\e_1, ... \, , \e_n)$ are not empirically well-defined and have shown that the empirically well-defined references to qualia are precisely the equivalence classes~\eqref{Intersub}. We now repeat a similar analysis for pre-models of consciousness. We first introduce the necessary mathematical tools.

Let $s \in \Aut(\E)$ be an element of the automorphism group~\eqref{AutGroup} which describes the freedom of an experiencing subject to choose labels for the qualia he/she experiences. Given a solution $(\bar \e_t, p_t)_t \in \D$, we may apply $s$ to that experience space in~\eqref{StateSpace} which is associated to the $i$th experiencing subject. This gives another trajectory
\beq\label{AutToFam}
	\big(\e_t^1, ... \, , s(\e_t^i), ... \, , \e_t^n,p_t\big)_{t \in \I}
\eeq
where $i \in \{1, ... \, , n\}$.
\noindent The map which takes $(\bar \e_t, p_t)_t$ to~\eqref{AutToFam} is an $\Aut(\E)$-action $\phi_i$ on $\K$, defined as
\begin{align}\label{Def_Phi_i}\begin{split}
	&\phi_i: \Aut(\E) \times \K \longrightarrow \K \\
	&\big(s, \big(\e_t^1, ... \, , \e_t^i, ... \, , \e_t^n,p_t\big)_{t} \big) \mapsto \big(\e_t^1, ... \, , s(\e_t^i), ... \, , \e_t^n,p_t\big)_{t} \,
\end{split}\end{align}
where the subscript $i$ indicates on which experience space $\Aut(\E)$ acts. We may take into account the freedom of every
experiencing subject to relabel his/her qualia by considering an action $\phi$ of 
\beq
\Autk := \Aut(\E) \times ... \, \times \Aut(\E)
\eeq
on $\K$, defined as
\begin{align}\begin{split}\label{Def_Phi}
	&\phi: \Autk \times \K \longrightarrow \K \\
	&\big(s_1, ...\, , s_n, \big(\e_t^1, ... \, , \e_t^i, ... \, , \e_t^n,p_t\big)_{t} \big) \mapsto \big(s_1(\e_t^1), ...\, , s_n(\e_t^n),p_t\big)_{t} \, .
\end{split}\end{align}
This action corresponds to the transformations we have considered in Section~\ref{RelabelingTrans}.
However, in the context of models of consciousness, this is not the most general form of relabelling. The most general form is
\begin{align}\begin{split}\label{Def_phi}
	&\sigma: \Autk \times \K \longrightarrow \K \\
	&\big(\bars, \big(\e^1_t,  ... \, , \e_t^n,p_t\big)_{t}\big) \mapsto \big(s_1(\e_t^1), ...\, , s_n(\e_t^n), p_t'\big)_{t} \: ,
\end{split}\end{align}
\ch{where similarly to~\eqref{Fam} we have used the shorthand $\bars := (s_1, ...\, s_n)$,} and where $p_t'$ is given by an action $\sigmaP$ of $\Autk$ on $\K|_P$,
\begin{align}\begin{split}\label{Def_varphi}
	&\sigmaP: \Autk \times \K|_P \rightarrow \K|_P\\
	&	\big(\bars, (p_t)_t \big) \mapsto (p_t')_t \: .
\end{split}\end{align}
This action~$\sigma$ reduces to the action~\eqref{Def_Phi} if~$\sigmaP$ is trivial. 
If~$\sigmaP$ is non-trivial, $\sigma$ specifies that the physical states are relabelled along with the qualia.
We will see below (cf. Section~\ref{Comparison} for details) that the possibility of a non-trivial~$\sigmaP$ is
what allows us to go beyond the standard methodology explained in Section~\ref{SecQualiaRef}.\smallskip

\noindent {\em Notation:} In what follows, we will use the shorthand $\bars (\bar \e_t) := (s_1(\e_t^1), ...\, , s_n(\e_t^n))$. As usual, we denote $\sigma \big(\bars, (\bar \e_t, p_t)_t \big)$ as $\sigma_\bars \big( (\bar \e_t, p_t)_t \big)$. Furthermore, we use $k:= \big(\bar \e_t,p_t\big)_{t} \in \K$.

\begin{Remark}\label{RemNecessary}\em
We remark that each action~$\sigma$ of the form~\eqref{Def_phi} has two different meanings:
{On the one hand}, they describe a relabelling of the trajectory $k$. I.e., $\sigma_{\bars}(k)$ describes the same situation
as~$k$ but with respect to a different choice of labelling. This is the meaning we have considered in Section~\ref{PhenomenologicalGrounding}. It is analogous to a change of reference frame in physics. {On the other hand}, $k':=\sigma_\bars(k)$ is simply another trajectory in $\K$, which for $\bars \neq \id \in \Autk$ describes a scenario which is genuinely different to that of~$k$. Whereas according to $k$, at time $t$ experiencing subject~$i$ experiences the quale he/she has labelled as $\e^i_t$ and physical state~$p_t$ pertains, according to $k'$ the experiencing subject experiences a quale he/she has labelled~$s_i(\e^i_t)$
and physical state $p_t'$ pertains. This is reminiscent of the distinction between active and passive transformations in physics. Using this terminology we have:
\begin{enumerate}[label=\arabic*.,topsep=0pt]
\item {\em Passive meaning of $\sigma$}: $k$ and $\sigma_\bars(k)$ are the {\em same} trajectory expressed in {\em different} labelling.
\item {\em Active meaning of $\sigma$}: $k$ and $\sigma_\bars(k)$ are {\em different} trajectories expressed in the {\em same} labelling.
\end{enumerate}
The fact that active and passive transformation have an identical mathematical form is related to the fact that qualia in virtue of their non-collatability cannot be referenced
intersubjectively.~\QEDrem \end{Remark}

Next, we use the fact that~$k$ and~$\sigma_{\bars}(k)$ describe the same trajectory with respect to two different choices of labels. 
Since a different choice of labels must not make a difference, it follows that if $k$ is a solution of~$M$, $\sigma_{\bars}(k)$ needs to be a solution as well, for any choice of $\bars \in \Autk$. This leads us to the following definition:

\begin{Def}\label{DefNecessary}\em
A {\em necessary condition for a pre-model of consciousness~$M$ to be empirically well-defined} is that there is an $\Autk$ action~\eqref{Def_phi} on~$\K$ which maps solutions to solutions, i.e. which satisfies
\beq\label{PhiSol}
	\sigma_\bars ( \D ) = \D 
\eeq
for all $\bars \in \Autk$.\footnote{Here, $\D$ is the set of solutions of~$M$ introduced above. Note that~\eqref{PhiSol} states an identity of sets. It is equivalent to $\sigma_\bars(k) \in \D$ for all $k \in \D$.}
\end{Def}

Using Definition~\ref{DefSym}, this yields the following lemma.

\begin{Lemma}\em\label{Necessary}
A necessary condition for a pre-model of consciousness~$M$ to be empirically well-defined is that
$\Autk$ is a symmetry group of~$M$ 
whose action is of the form~\eqref{Def_phi}.
\end{Lemma}

\Proof
According to Definition~\ref{DefSym}, $\Autk$ is a symmetry of the model~$M$ iff~\eqref{Def_phi} is effective and leaves~$\D$ invariant.
Invariance holds by Definition~\ref{DefNecessary}. Effectivity holds because for large enough~$\K$ (cf. Footnote~\ref{Regularity})
every action of the form~\eqref{Def_phi} is effective: For any $\bars \in \Autk$ with $\bars \neq \id$, there exists an $\e^i_t \in \E$ such that
$s_i (\e^i_t) \neq \e^i_t$ as well as a trajectory~$k \in \K$ which contains this label, so that $\sigma_\bars(k) \neq k$.
\QED

If there \ch{are} only collatable aspects of experience, the automorphism group $\Aut(\E)$ is trivial, so that~\eqref{PhiSol} is satisfied. Therefore, in this case, all pre-models of consciousness are empirically well-defined. If there are non-collatable aspects of experience, however, $\Aut(\E)$ is non-trivial, so that~\eqref{PhiSol} poses a condition that needs to be satisfied, and Lemma~\ref{Necessary} shows that the condition is in fact that there is an $\Autk$ symmetry.

Thus Lemma~\ref{Necessary} establishes the mathematical consequences of non-collatability: The need of an $\Autk$ symmetry in a model of consciousness. Since the existence of a symmetry is dependent on the dynamical trajectories of a model of consciousness, which are in turn determined by its laws~$\L$, the condition posed by the lemma is in fact a requirement with respect to the model's laws.

\subsection{The Mathematical Structure of Models of Consciousness}\label{SecDefMoc}
In the previous section, we have found that the existence of an action~\eqref{Def_phi} which constitutes a symmetry 
is a necessary condition for a pre-model of consciousness to be empirically well-defined. We therefore need to include this requirement when specifying what constitutes a general model of consciousness. The result is given in the following definition. It specifies the necessary structure of any model of consciousness which is to address any non-collatable aspects of experience, e.g.
ineffable, private or inaccessible aspects. In particular, any model which aims to address any aspect of experience
which is referenced by the Nagelian ``what it is like'' conception (Remark~\ref{ExWhatIsItLike}) necessarily needs to carry this mathematical structure.

\begin{Def}\em \label{DefModel}
A {\em model of consciousness} is a pre-model of consciousness~$M$ as defined in Definition~\ref{DefPreModel} which additionally carries an~$\Autk$ symmetry of the form~\eqref{Def_phi}.
\end{Def}

We remark again that the additional requirement relate to the laws~$\L$ of a pre-model of consciousness~$M$; the laws need to be such that there is an action~$\sigmaP$ which turns~\eqref{Def_phi} into a symmetry.\smallskip

In Section~\ref{Comparison}, we will show that this framework indeed allows \ch{us} to go beyond the limitations of the standard approach explained in Section~\ref{SecQualiaRef}.
To furthermore make the point that this is a sufficient mathematical framework for the scientific study consciousness (i.e. sufficient to study qualia proper, cf. Definition~\ref{DefTask}), we show in the following example that a typical class of ideas put forward in the neuroscience of consciousness can indeed be formalized in this framework: the idea that qualia are determined by physical states.

\begin{ExampleS}\em\label{ExamplePdeterminesL}
We consider the hypothesis that qualia are \emph{determined} by physical states, e.g. by neural activity in the brain. What exactly constitutes the determination does not matter in what follows. Examples are type identity theory, where qualia are types of physical states, or functionalism, where qualia are functional roles which are determined by physical states. Another example is non-reductive functionalism~\cite{chalmers1995absent} where the physical states determine functional roles, which in turn determine qualia via non-reductive laws of nature.

 The naive formalization of this idea would be to consider a function (in the mathematical sense) which specifies which quale an experiencing subject experiences for each physical state $p \in P$. As we have seen in Section~\ref{PhenomenologicalGrounding}, the problem with this naive formalization is that qualia cannot be referenced intersubjectively, so that the specification of a function of this type is only meaningful up to the equivalence~\eqref{AutEquiv}.  In order to properly formalize this idea, we proceed as follows. 

For simplicity, we consider the case of one experiencing subject ($n=1$).
We assume that a particular labelling has been fixed by the experiencing subject and assume that with respect to this labelling a function
\begin{align}\label{ExDetermineF}
f: P \rightarrow \E \qquad p \mapsto f(p)
\end{align}
is given, where $P$ is the state space of a physical theory~$T_P$ as above and $\E$ denotes the experience space.
This function expresses in which way qualia are determined by physical states and could, ideally, be the result of experiments which include the experiencing subject in question. The state space~$P$ could, e.g., refer to neural activity. 

Based on this function~$f$, we can define a pre-model of consciousness~$M$. To this end, we set $d= \E \times P$, choose $\K$ as the right hand side of~\eqref{defK} and define the solutions of the model in terms of the solutions $(p_t)_t \in \D_P$ of the physical theory $T_P$ as
\begin{align}\label{ExDetermineD}
\D = \big\{ (f(p_t),p_t)_t \, \big| \, (p_t)_t \in \D_P \big\} \:.
\end{align}
The solutions of this model are thus given by the solutions of the physical theory (e.g. brain dynamics) equipped with qualia as specified by~$f$.

As it stands, this model is {\em not} invariant with respect to relabelling. E.g., if the choice of labels is being changed
according to some $s \in \Aut(\E)$, the solution $(f(p_t),p_t)_t $ is being mapped to the solution $(s(f(p_t)),p_t)_t $ which in general will not be an element of $\D$ as defined in~\eqref{ExDetermineD}. Thus the theory is not empirically well-defined. 

In order to establish empirical well-definedness, there are two choices:
First, one could demand that $s(f(p)) = f(p)$ for all $s \in \Aut(\E)$ and all $p \in P$. This amounts to considering a function $f: P \rightarrow \E \backslash\!\!\! \sim$, where $ \E \backslash\!\!\! \sim$ is as in~\eqref{Intersub}, which does not achieve the task of Definition~\ref{DefTask}. The alternative is to specify an action~$\sigmaP$ as in~\eqref{Def_varphi}, as we now explain.

The action~$\sigmaP$ describes how the physical state changes along with a change of qualia (active interpretation, cf. Remark~\ref{RemNecessary}). We observe that a definition of~$\sigmaP$ as 
\beq\label{ExDetermineSigmaP}
\sigmaP_s \big( (p_t)_t \big) := \big(p_t'\big)_t \qquad \textrm{ with } \qquad p_t' :=  f^{-1}_\cdot \big(s(f(p_t) ) \big) \:,
\eeq
where $f^{-1}_\cdot(\e)$ denotes any element of the pre-image $f^{-1}(\e)$ of $\e$,
yields for~\eqref{Def_phi}
\begin{align*}
\sigma_s\big( (f(p_t),p_t)_t \big) = \big( s(f(p_t)),p'_t \big)_t = \big( f(p'_t),p'_t \big)_t \:,
\end{align*}
where we have used $f(p_t') = f \circ f^{-1}_\cdot \big(s( f(p_t) ) \big) = s( f(p_t) )$. Thus if $(p_t')_t$ is a solution of~$T_P$, the action~\eqref{Def_phi} with~$\sigmaP$ as defined in~\eqref{ExDetermineSigmaP} is a symmetry of~$M$, so that~$M$ is a model of consciousness, i.e. empirically well-defined.

Thus the idea that physical states determine qualia can indeed be formalized, even though qualia are defined to be non-collatable aspects of experience. The limitation of this approach is that the function~$f$, being defined with respect to a particular choice of labelling of the experiencing subject, cannot be interpreted as specifying {\em the} quale which the experiencing subject experiences along with a particular physical state~$p$. Nevertheless, the formalism allows us to treat the case that a quale, whichever one it is among the qualia in the equivalence class~$[f(q)]$, is determined by the physical state~$p$. Here, the equivalence class in question is the one defined in~\eqref{AutEquiv}.
A further analysis of the difference to a direct description will be given in Section~\ref{Comparison}.
\end{ExampleS}

We close this section by specifying the empirically well-defined part of the trajectories of a model of consciousness~$M$. This specification is analogous to the specification of empirically well-defined sequences in~\eqref{EqClassC}.

As in Section~\ref{PhenomenologicalGrounding}, we define two trajectories $(\bar \e_t, p_t)_t$ and $(\bar \e_t', p_t')_t \in \K$ to be equivalent if one can be obtained from the other by relabelling the qualia of the experiencing subjects. 
In contrast to Section~\ref{PhenomenologicalGrounding}, relabelling is defined in terms of the action~\eqref{Def_phi}, which for non-trivial~$\sigmaP$ includes a relabelling of the physical states. We denote this equivalence by $\sim_\sigma$,
\begin{align}\label{Equiv}\begin{split}
(\bar \e_t, p_t)_t \sim_\sigma (\bar \e_t', p_t')_t \quad &\textrm{ if and only if there is an } \bars \in \Autk \\
&\textrm{ such that }  \big(\bar \e_t', p_t'\big)_t = \sigma_\bars\big( (\bar \e_t, p_t)_t \big)  \: .
\end{split}\end{align}
Note that $\sigmaP$, and hence $\sigma$, depends on the laws of the model~$M$ under consideration.
The empirically well-defined part of the trajectories is given by the quotient set
\beq\label{IntersubMeaningful}
	\bigslant{\K}{\sim_{\tiny \sigma}} \:,
\eeq
i.e. by the the space of equivalence classes of~$\sim_\sigma$.
This space describes the distinctions which remain once all trajectories are identified which can be mapped to each other by relabelling the qualia of the experiencing subjects.\smallskip

\subsection{Comparison with Direct Reference}\label{Comparison}
In the previous sections, we have shown that non-collatability implies that models of consciousness need to carry a particular symmetry group in order to be well-defined and that this allows \ch{us} to address qualia proper, i.e. individual non-collatable aspects of experience. In this section, we compare the methodology so introduced with what may be called a `direct reference' of qualia:
A description of experimental data or theoretical idea simply in terms of qualia's labels, without invoking any of the mathematical details introduced in Definitions~\ref{DefPreModel} and~\ref{DefModel}.

Mathematically, a direct reference of the qualia of~$n$ experiencing subject is simply a family
\begin{align}\label{CompFam}
\big(\e^1_t, ... \, , \e^n_t\big)_{t \in \I} \: ,
\end{align}
where $\I$ is some parameter space as above. For example, this could a time-series of reports of experiencing subjects.
Whereas a direct description may ignore the mathematical details of Definitions~\ref{DefPreModel} and~\ref{DefModel}, it cannot ignore the ambiguity induced by non-collatabiltiy of lables, which is fundamental and independent of any of the mathematical tools introduced in Sections~\ref{SecMathStrMoc} and~\ref{PhenGrMethod}.

In Section~\ref{SecQualiaRef}, we have studied this ambiguity and what it implies for references to qualia in detail, and found that the corresponding well-defined statement is given by~\eqref{EqClassC}. In the present notation, this reads
\begin{align}\label{CompEquiv}\begin{split}
	(\bar \e_t)_t \sim (\bar \e_t')_t \quad &\textrm{ if and only if there is an } \bars \in \Autk\\
	&\textrm{ such that } {\e_t'}^i = s_i(\e^i_t) \textrm{ for all } t \in \I 
\end{split}\end{align}
and
\begin{align}\begin{split}\label{CompEqClassC}
\big[(\bar \e_t)_t \big] :=& \big\{ (\bar \e_t')_t \, \big| \, (\bar \e_t')_t \sim (\bar \e_t)_t \big \} \: .
\end{split}\end{align}
In order to compare this with a model of consciousness, we assume that one wishes to relate (e.g. statistically analyse) the direct reference of qualia with some properties of a physical system, e.g. neural activation patterns. We assume that these properties are determined by states of the physical system and denote the \ch{state} space as above by~$P$.
Thus the data under consideration is of the form
\begin{align}\label{CompData}
\big(\e^1_t, ... \, , \e^n_t,p_t\big)_{t \in \I} \: .
\end{align}
It could result, e.g., from of verbal reports of experience and simultaneous fMRI scans or EEG recordings.

In~\eqref{IntersubMeaningful}, we have found that the empirically well-defined trajectories of a model of consciousness are given by the quotient set 
\beq\label{CompIntersubMeaningful}
	\bigslant{\K}{\sim_{\tiny \sigma}} \:.
\eeq
The following lemma gives the corresponding result for the direct description.
\begin{Lemma}\em\label{LemmaDirectIntersubMeaning}
The {\em empirically well-defined} trajectories of a {\em direct description} of qualia are given by the quotient set
\beq\label{DirectIntersubMeaningful}
	\bigslant{\K}{\sim_{\tiny \phi}} \:,
\eeq
where $\phi$ is the action~\eqref{Def_Phi} and where $\sim_{\tiny \phi}$ is defined as in~\eqref{Equiv}.
\end{Lemma}

This lemma states what a direct description of qualia can reference in light of non-collatability. It expresses the epistemic constraints which {ineffability}, {privateness}, {inaccessibility} and other characteristic features which imply non-collatability pose for any theoretical or experimental account of consciousness.

The difference between~\eqref{CompIntersubMeaningful} and~\eqref{DirectIntersubMeaningful} is that in~\eqref{CompIntersubMeaningful} one takes the quotient with respect to an action that generally acts non-trivially on the physical trajectories~$(p_t)_t$, whereas in~\eqref{DirectIntersubMeaningful}
one takes the quotient with respect to an action that acts trivially on the latter.
We defer further discussion to after the proof.
\Proof[Proof of Lemma~\ref{LemmaDirectIntersubMeaning}]
In terms of the notation~\eqref{CompData}, the equivalence~\eqref{CompEquiv} is given by
\begin{align}\label{CompEquivData}\begin{split}
	(\bar \e_t,p_t)_t \sim (\bar \e_t',p_t')_t \quad &\textrm{ if and only if there is an } \bars \in \Autk\\
	&\textrm{ such that }  (\bar \e_t',p_t')_t = \phi_\bars \big( ( \bar \e_t,p_t)_t \big) \:.
\end{split}\end{align}
According to~\eqref{Equiv}, this is precisely the equivalence~$\sim_\phi$. Hence the empirically
well-defined trajectories~\eqref{CompEqClassC} are elements of the quotient set~\eqref{DirectIntersubMeaningful}.
\QED

The quotient space~\eqref{DirectIntersubMeaningful} contains the empirically well-defined trajectories that can be
referenced by a direct description of qualia in light of non-collatability. The quotient space~\eqref{CompIntersubMeaningful}, on the other hand,  gives the empirically well-defined trajectories of a model of consciousness. The difference between both is determined by the action~$\sigmaP$ defined in~\eqref{Def_varphi}, which describes how the physical states transform if one relabels the qualia of an experiencing subject (passive meaning, cf. Remark~\ref{RemNecessary}), but also how the physical states change if the qualia of an experiencing subject change (active meaning in Remark~\ref{RemNecessary}). It is precisely the possibility of a non-trivial~$\sigmaP$ which allows the laws postulated by a model of consciousness to address individual qualia to a certain extent. The \ch{extent} to which this is possible is limited by the requirement that~\eqref{Def_phi} constitutes a symmetry of the Model~$M$.

Mathematically, this is reflected in the fact that elements of~\eqref{DirectIntersubMeaningful} are of the form
\begin{align}\label{CompFormDirect}
\big( [ (\e^1_t)_t ] , ... \, , [ (\e^n_t)_t ], (p_t)_t \big) \:,
\end{align}
where each equivalence class $[ (\e^i_t)_t ]$ is as in~\eqref{CompEquiv} and~\eqref{CompEqClassC}.
This coincides with our result in Section~\ref{RelabelingTrans} on references to qualia (cf. quotient space~\eqref{Intersub}).
The elements of~\eqref{CompIntersubMeaningful}, on the other hand, are not of this form: A non-trivial $\sigmaP$
results in equivalence classes in which the physical states are {\em mixed} with the labels of the various experiencing subjects.  The empirically well-defined trajectories cannot be separated as in~\eqref{CompFormDirect}.

This means that if one chooses a direct description of qualia and investigates the relation to the physical domain, one first has to consider equivalence classes~\eqref{CompEqClassC} (for only they are empirically well-defined) and subsequently propose or analyse the relation to the physical domain. Using a model of consciousness, one may exchange this order: One may first postulate a relation of qualia and the physical domain, and subsequently remove the arbitrariness of choosing labels by considering equivalence classes~\eqref{Equiv}. The requirement of there being a symmetry $\sigma$ simply makes sure that this relation (the law~$\L$ of a model of consciousness) is chosen in such a way that the second step -- obtaining an empirically well-defined theory -- is possible at all.
A non-trivial~$\sigmaP$ implies that after the second step, one does not end up with what one could have obtained using the first procedure right away. This small detail can have large empirical consequences.
\smallskip

Thus we have shown that models of consciousness are more powerful than direct references in regard to the scientific study of non-collatable aspects of experience. Since collatable aspects are always contained in our formalization as a special case (trivial $\Aut(\E)$), we see that mathematical models of consciousness provide a suitable methodology for the scientific study of consciousness which allows \ch{us} to take the key characteristics of {ineffability}, {privateness} and {cognitive, linguistic and communicative inaccessibility} into account.

We conclude this section by remarking that Lemma~\ref{LemmaDirectIntersubMeaning} shows that if~$\sigmaP$ is chosen as trivial in Definition~\ref{DefModel}, the result is a direct description of qualia as referenced here. Thus direct references constitute a special type of model of consciousness, and models of consciousness are in fact a genuine generalization of direct references.

\section{Closure of the Physical}\label{ClosureOfPhys}

Great care has been taken in the previous sections to motivate all constructions in an operational, epistemological or phenomenological way, making sure that they are independent of any metaphysical commitment. \ch{Metaphysical choices should only be made when constructing individual models of consciousness.}
In this section, we make a brief remark about a particularly important metaphysical choice, the closure of the physical.

The closure of the physical is often called ``causal closure of the physical''~\cite{Bishop.2005}. It denotes the idea that ``physical laws already form a closed system''~\cite[p.\,17]{Chalmers.2010} and is an important assumption which underlies many philosophical and scientific investigations of consciousness.
We mention it here because, as explained in Appendix~\ref{ChalmersGrounding}, Chalmers' grounding of the scientific study of consciousness needs to make use of the closure of the physical in its definitions, which limits the applicability of this grounding substantially (cf. Appendix~\ref{APIssuesChalmers}). 

This is different for the grounding put forward in Section~\ref{NewGrounding}. Since none of the basic definitions or formal constructions refers to the closure of the physical in any way, this is in fact an independent assumption which one may or may not make when constructing models of consciousness. As a result, the conceptual and formal frameworks developed here are suitable also to construct models of consciousness which express metaphysical ideas such as dual aspect monism or property dualism that do not describe the physical as closed.

In terms of the formalism developed in Sections~\ref{SecMathStrMoc} and~\ref{PhenGrMethod}, the assumption of the closure of the physical 
can be stated in a particularly concise form.

\begin{Def}\em \label{DefClosurePhys}
A model of consciousness~$M$ describes {\em the physical as closed}
if and only if
\beq\label{ClosurePhysEq}
	\D|_P = \D_P \:,
\eeq
where~$\D|_P$ is defined in~\eqref{D-P} and where $\D_P$ has been introduced in Section~\ref{Theories} to denote the solutions of the physical theory~$T_P$ which underlies the model~$M$.
\end{Def}

 This definition says that a model of consciousness~$M$ describes the physical as closed if and only if the physical trajectories which are determined by the
laws of~$M$ (as part of the solutions~$\D$)
are, as a set, equal to the solutions of the physical theory~$T_P$ which~$M$ is based on. Whether or not~\eqref{ClosurePhysEq} is satisfied depends on the laws~$\L$ postulated by a particular model.

This concludes our brief excursion to metaphysics. In the next section, we review several examples and how they relate to the formalism introduced here. A conceptual point about the closure of the physical is made in Appendix~\ref{AppendixClosure}.

\section{Examples}\label{Examples}

In this section, we review some models of consciousness that have been proposed in the literature and explain how they relate to the formalism introduced in Sections~\ref{SecMathStrMoc} and~\ref{PhenGrMethod}, and to the concepts introduced in Section~\ref{NewGrounding}.

\subsection{Integrated Information Theory}\label{ExIIT}
Our first example is Integrated Information Theory (IIT) which has been proposed by Giulio Tononi in 2004~\cite{Tononi.2008}
and has since been developed considerably. The current version of that theory~\cite{IIT30} consists of an algorithm whose input is a model of a physical system (together with a state of that system and including its dynamical laws) and whose output are formal quantities which give answers to the following three questions: 1.~Which parts of the system are conscious? 2.~What are they conscious of? 3.~How conscious are they?

To answer the first question, the theory's algorithm identifies some (mutually disjoint) subsystems of the system.
To answer the second question, for each such subsystem~$S$, the algorithm specifies what is called a `maximally irreducible conceptual structure' (MICS). This is a mathematical object of the following kind: Let $\mathcal P_S$ be the space of probability distributions (or probability measures) over the states of the subsystem~$S$. A `concept' is an element of the space\footnote{In the terminology of~\cite{IIT30}, a concept consists of the maximally-irreducible cause and effect repertoires of a mechanism~$M$ of~$S$ together with its integrated information
$\varphi(M)$, provided that the latter is non-zero~\cite[Supplementary S1, p.\,176]{Mayner.2018}.}
\[
	\mathcal P_{CS}:=\mathcal P_S \times \mathcal P_S \times \R_0^{+} \: .
\]
The maximally irreducible conceptual structure is an $n$-tuple of concepts, where $n$ is determined dynamically by the theory and may vary from subsystem to subsystem. I.e., it is an element of the ``qualia space''~\cite[graphical illustration in Figure\,15]{IIT30}
\beq\label{IITQualiaSpace}
	\E_S :=   \mathcal P_{CS}^{\times n(S)} \:.
\eeq
Finally, in order to answer the third question, the algorithm specifies the integrated conceptual information $\Phi^{\textrm{max}}(S) \in \R^+_0$. In summary:
\begin{quote}
``[T]he central identity [of IIT] is the following: The maximally irreducible conceptual structure (MICS) generated by a [subsystem $S$] is identical to its experience. The
constellation of concepts of the MICS completely specifies the quality of the experience (...). Its irreducibility $\Phi^{\textrm{max}}$ specifies its quantity.''~\cite[p.\,3]{IIT30}.
\end{quote}

The main papers of the theory remain somewhat silent about what exactly they take the terms ``consciousness'' 
or ``quality and quantity of an experience''~\cite{IIT30} to mean. The first observation is that ineffable aspects of conscious experience seem to have played at least a small role in the early development of the theory. E.g.,
in~\cite[p.\,229]{Tononi.2008}, Tononi notes that ``[t]he notions just sketched aim at providing a framework
for translating the {\em seemingly ineffable} qualitative properties 
of phenomenology into the language of mathematics'' (our emphasis).
As we have explained in detail in Section~\ref{PhenomenologicalGrounding}, ineffable aspects of experience cannot be put in a one-to-one correspondence with mathematical objects, simply because two or more experiencing subjects have no means to ensure that they have associated the same ineffable aspect of experience with the same mathematical object. This was the reason for us to introduce experience spaces~$\E$ via labels in Section~\ref{PhenomenologicalGrounding} and what lead to the requirement of there being a symmetry that describes relabelling.
Following this path, we might take IIT's ``qualia space''~$\E_S$ to constitute the experience space of qualia as
defined in Definition~\ref{DefQualia}.

This brings us to the question of how collatable relations between qualia so defined, such as the ones put forward in Phenomenological \ch{Axiom}~\ref{PFcollatable}, are related to the mathematical structure of the space~$\E_S$. 
First, we note that the space $\E_S$ can be equipped with a metric: For any metric $d$ on $\mathcal P_S$ and using the usual metric on $\R^+_0$, summation allows us to define both a metric on~$\mathcal P_{CS}$ and $\E_S$. This metric can be taken to express the similarity relation in Phenomenological \ch{Axiom}~\ref{PFcollatable}. And indeed, this may again have been a guiding idea in the development of the theory, ``experiences are similar if their shape is similar''~\cite[p.\,228]{Tononi.2008}. 
Next, what has been called the ``intensity'' of an experience in Example~\ref{ExamplePFRel} corresponds to the ``quantity'' of experience according to IIT. The corresponding mathematical structure is the $\R^+_0$ in which $\Phi^{\textrm{max}}(S)$ takes values.

IIT arguably encodes another collatable relation between qualia: Their {\em composition} in experience.
This is usually formulated as an axiom of IIT which states that ``[c]onsciousness is compositional (structured): each experience consists of multiple aspects in various
combinations''~\cite[p.\,2]{IIT30}. 
The composition of the experience of a subsystem~$S$ in terms of more elementary experiences of the same subsystem 
is modelled by the Cartesian product structure of~\eqref{IITQualiaSpace} in terms of the concept spaces~$\mathcal P_{CS}$.
One may interpret the $\R^+_0$ that constitutes the last factor of~$\mathcal P_{CS}$ as the intensity of the the more elementary experiences.\smallskip

In summary, the basic definitions of IIT seem to fit quite well with the basic definition of the phenomenological grounding 
put forward in Section~\ref{SecPhenGr} and IIT
can be taken to constitute a pre-model of consciousness as defined in Definition~\ref{DefPreModel}.
However, in order to take into account the non-collatability of the corresponding aspects of experience,
the symmetry~\eqref{Def_phi} has to be implemented. This can be done by simply swapping the states of the physical system that give rise to particular labels $\e \in \E_S$: If $\e_1, ... \, \e_n \in \E_S$ are the labels of the conscious subsystems of a system in state~$p_1$, and $\e'_1, ... \, \e'_{n'} \in \E_S$ are the labels of conscious subsystems of the system in state~$p_2$, for any $\bars \in \Aut(\E_S)$ which maps the former to the latter, we define the action~\eqref{Def_varphi} as $\sigmaP_\bars (p_1) = p_2$. (For details, cf. Example~\ref{ExamplePdeterminesL}.) Equipped with this symmetry, IIT constitutes a model of consciousness as defined in Definition~\ref{DefModel}.
\smallskip

We conclude this example with some conceptual remarks. First of all, we note that the phenomenological grounding approaches model-building differently than~\cite{IIT30}. Whereas in the latter, phenomenological \ch{axioms} are used to justify the definition of the algorithm, i.e. the dynamical equations of IIT, the phenomenological grounding uses phenomenological \ch{axioms} to model the mathematical space associated with qualia. We have seen above that an earlier version of IIT put forward in~\cite{Tononi.2008} is more aligned with this perspective.
From the perspective of the phenomenological grounding, the main task would be to motivate the mathematical structure of~\eqref{IITQualiaSpace} in more detail. E.g., one could ask why the elementary qualia are labelled by elements of the space $\mathcal P_S \times \mathcal P_S$ and not by a simpler metric space? Does the structure of the former have any phenomenological interpretation? Questions of this sort may have large consequences for the further development of the theory because the algorithm of IIT in its current form makes essential use of elements of $\mathcal P_S \times \mathcal P_S$.

Second, we remark that IIT describes the physical as closed: The dynamical evolution of the physical domain is not changed in any way by the theory. Thus, if one interprets the theory in Chalmers' or the phenomenological grounding (or similar ones, in fact\footnote{%
This problem seems to appear in any grounding which exhibits an explanatory gap. One could try to avoid the problem by interpreting~\eqref{IITQualiaSpace} in terms of aspects of experience that do not exhibit an explanatory gap.
This, however, would raise the question of why a novel law (the algorithm of IIT) should determine those aspects, as compared to some form of neural processing.
})
the question arises of how the theory's mathematical postulates -- first and foremost the algorithm it specifies -- can be evaluated experimentally at all, for it seems that all results one can hope to obtain from neuroscientific experiments that scan the brain (EEG, fMRI, etc)
are physical and therefore determined by the physical domain alone. Put in simple terms, one may ask what one actually learns when collecting experimental data that is in fact completely determined by the physical domain. This argument is outlined in more detail in Appendix~\ref{AppendixExperiments} and related to the transcendental argument against the closure of the physical given in Remark~\ref{NoClosureOfPhys}. In Section~\ref{ExIITCollapse}, we review a modification of IIT which avoids this problem.

\subsubsection{Integrated Information-Induced Quantum Collapse}\label{ExIITCollapse}
To overcome the problem mentioned in the last paragraph, one needs to propose a model of consciousness which does not postulate the physical as closed. 
A first model of this kind based on Integrated Information Theory (IIT) is given in~\cite{KremnizerRanchin.2015}.
It refers to an early version of IIT which only answers the third question of Section~\ref{ExIIT}: How conscious is a physical system?

The physical theory on which this model is based is a quantum system with Hilbert space~$\H$ and Hamiltonian $H$. The states are density matrices $\rho \in L(\H)$. Given a density matrix~$\rho$, the {\em Quantum Integrated Information} is defined as
\begin{align}\label{QIIT}
\Phi(\rho) = \inf \Big\{ S\big(\, \rho \, \big|\!\big| \, \otimes_{i=1}^N \Tr_i\rho\, \big) \Big\} \:,
\end{align}
where the infimum is taken over all decompositions of the Hilbert space~$\H$, i.e. over all isomorphisms between~$\H$ and 
a Hilbert space of the form~$\H_1 \otimes ... \otimes \H_N$, where $\Tr_i\rho$ denotes the reduced density matrix on the Hilbert space~$\H_i$ (i.e.~$\Tr_i$ is a trace over all $\H_j$ with $j \neq i$) and where finally $S(\rho \| \rho')$ denotes the quantum relative entropy defined as
\begin{align*}
S(\rho \| \rho') = \Tr \rho \log \rho - \Tr \rho \log \rho' \: .
\end{align*}
Here,~$\Tr$ denotes the trace over the whole Hilbert space~$\H$.
According to this model,~\eqref{QIIT} specifies how conscious the physical system is in the state~$\rho$.

In order to specify how consciousness in turn influences the physical domain, the model modifies the time-evolution of the physical system. Whereas in quantum theory, the time-evolution of a closed system with Hamiltonian~$H$ is determined by
\begin{align*}
\frac{\partial \rho}{\partial t}   =- \frac{i}{\hbar} [H,\rho] \:,
\end{align*}
this models proposes the evolution equation
\begin{align*}
\frac{\partial \rho}{\partial t}   =- \frac{i}{\hbar} [H,\rho] + \sum_{n,m=1}^{N^2-1} h_{n,m}\big( \Phi(\rho) \big) \cdot \big( L_n \rho L_m^\dagger - \tfrac{1}{2} \, \rho L_m^\dagger L_n + \tfrac{1}{2} \, L_m^\dagger L_n \rho \big)
\:,
\end{align*}
where $ h_{n,m}( \Phi(\rho) )$ are continuous functions of~$\Phi(\rho)$ which vanish if $\Phi(\rho) = 0$
and where~$L_k$ are operators on~$H$. 
This is a Lindblad evolution equation which describes, among other things, models of spontaneous wave function collapse. 
By choosing the functions~$h_{n,m}$ suitably small, one can make sure that the model is compatible with physical experiments to date. 
The model is furthermore experimentally accessible in that it predicts a collapse rate which is dependent on $\Phi(\rho)$, rather than mass or the number of particles alone, as is the case in other spontaneous collapse models (cf.~\cite[Sec.\,5]{KremnizerRanchin.2015}).

\subsection{Global Neuronal Workspace Theory}\label{SecGNW}
The Global Neuronal Workspace model (GNW) is, next to Integrated Information Theory, the other model largely favoured by neuroscientists.
In contrast to the latter, however, it is usually stated directly in terms of brain physiology~\cite{DehaeneChangeuxNaccache.2011,DehaeneKerszbergChangeux.1998}. Even though this is sufficient to make some specific predictions~\cite{DehaeneChangeuxNaccache.2011}, a more formal model would be desirable, not least to make a detailed comparison with IIT possible.

In what follows, we outline how a formal model could be constructed which takes as input any physical system (in a certain class of systems) and determines what the system is conscious of. To this end, we apply concepts from dynamical systems and nonbinary information processing whose connection with consciousness has recently been suggested in~\cite{Grindrod.2018}. While this attempt is {\em very} preliminary, the hope is that a genuine formal model can be developed along these lines in future work. A different goal is pursued in~\cite{Wallace.2005}.\smallskip

Let $S$ be a physical system. We assume that it consists of a set $N_v$ of components (`vertices', representing neurons in a neuronal network), each of which is in a particular state $u_i(t)$. Here, $t \in \I$ denotes time and $i \in N_v$ denotes the component in question. Furthermore, we assume that it consists of a set $N_e$ of directed edges (representing axons, dendrites, synapses, etc. in a neuronal network), each of which may be in a state $w_l(t)$, $l \in N_e$, (e.g. representing the synaptic strength in a neuronal network).
As usual, we define the parents $\Pa_i$ of the $i$th component to be those components from which a directed edge leads to $i$, and assume $i \in \Pa_i$. 
Finally, the dynamics of the system are specified component-wise by a set of `update-rules' $(f_i)_{i \in N_v}$, where $f_i$ determines how the state $u_i(t)$ of the $i$th component depends on the states of its parents and the states of the edges coming from its parents at previous times.

This system has conscious representations, according to the GNW model, if two necessary conditions are satisfied. The first of these is that the system has ``two main computational spaces, each characterized by a distinct pattern of connectivity''~\cite[p.\,56]{DehaeneChangeuxNaccache.2011}. The first computational space is a ``processing network, composed of a set of parallel, distributed and functionally specialized processors or modular subsystems subsumed by topologically distinct (...) domains with highly specific local or medium-range connections''~[ibid.]. The second computational space is a 
``a global neuronal workspace, consisting of a distributed set of (...) neurons characterized by their ability to receive from
and send back to homologous neurons in other (...) areas horizontal projections through long-range excitatory axons''~\cite[p.\,56]{DehaeneChangeuxNaccache.2011}.

In order to construct a formal model of consciousness based on this hypothesis, a definition has to be given which specifies which structure in a physical system
counts as a computational space of each kind, and which not; i.e. a definition of the necessary ``patterns of connectivity'' in terms of the mathematical structure of a physical system. In order to propose a such a definition, we combine the ideas of the GNW model with some of the ideas put forward in~\cite{Grindrod.2018}, using in particular the similarity between the processing network described by GNW and the  ``global directed network consisting of a large sparsely connected array of much smaller, irreducible subgraphs (ISGs), representing directed neuron-to-neuron connections'' put forward in connection with consciousness in~\cite{Grindrod.2018}. Here, an ISG is defined as follows.
\begin{enumerate}
\item[{[D1]}] A set $N_{\textrm{ISG}} \subset N_v$ of components of the physical system constitutes an irreducible subgraph (ISG) if there is a directed edge between any ordered pair of components in this set~\cite[p.\,25]{Grindrod.2018}.
\end{enumerate}
The similarity to the processing network of the GNW model comes about due to the major observations in~\cite{Grindrod.2018} that each ISG ``acts as an analog filter, a dynamical decision-maker (preferring one or another resonant mode), an amplifier, and a router''~\cite[p.\,27]{Grindrod.2018}.
In order to specify a necessary pattern for the global neuronal workspace, we simply require that this is a network with a directed edge going into and coming out of each ISG, noting that a further requirement on this network will be added below.
In summary, a proposal for the first necessary condition for the system $S$ to be conscious may be put as follows.
\begin{enumerate}
\item[{[N1]}] The system $S$ needs to contain two disjoint subsets $N_p,N_g \subset N_v$ of components: First, a set $N_p$ of components whose induced subnetwork is a network of ISGs, were the inter-ISG-connections are feed-forward only.
Second, a set $N_g$ of components with directed edges going from this set into all ISGs, and directed edges going to this set from all ISGs.
\end{enumerate}
Clearly, this definition is preliminary and will have to be improved substantially to \ch{facilitate} a full-fledged model to be defined.

In~\cite{Grindrod.2018}, general properties of the state of ISGs are explained, which have been found in previous work.
In particular, if the system satisfies a few conditions, including the feed-forward property mentioned in definition [N1], the ISGs will typically 
carry out ``successive pattern recognition tasks exploiting both remembered contextual information and prior expectations from past events
(...) as well as the assumption of the structures (elements) that are identified at [a] previous level''~\cite[p.\,30]{Grindrod.2018}.
The result of this task is recorded by a dynamical attractor on the ISG's components, which we denote by $m_k(t)$, where $k$ indexes the ISGs in the system.
These dynamical attractors represent ``perceived sources/objects, (...) events, (...) narratives, (...) scenarios''~\cite[Fig.\,2]{Grindrod.2018}. 
Combining these results with the idea that
``[t]he entire workspace is globally interconnected in such a way that only one such conscious representation can be active at any given time''~\cite[p.\,58]{DehaeneChangeuxNaccache.2011}, we arrive at a proposal for the second necessary condition for the system to be conscious:
\begin{enumerate}
\item[{[N2]}] The induced subnetwork of $N_g$ needs to be such that at any time $t$, its state `represents'
only one of the ISGs' dynamical attractors $m_k(t)$.
\end{enumerate}
Here, one could, e.g., define the term ``represent'' to mean that at any time $t$ the state of the network is (essentially) equal to one of the states $m_k(t)$, but other more realistic choices might be possible.

If both necessary conditions [N1] and [N2] are satisfied at a particular time $t$, the GNW model claims that the system $S$ is conscious of the ``perceived object, event, narrative or scenario'' $m_k(t)$ represented in the global workspace network $N_g$. Due to the directed edges from $N_g$ to the ISGs, 
the state of the ISG $k$ may be made ``directly available in its original format to all other workspace processes''~\cite[p.\,15]{DehaeneNaccache.2001}.\smallskip

Clearly, this outline leaves open various questions. Most notably, the question of how modes $m_k(t)$ of ISGs may relate to experience. Whereas IIT's qualia space~\eqref{IITQualiaSpace} has some structure which relates to phenomenology, it is highly questionable whether this can be asserted of the states of ISGs, which behave generically like a small number of ``monotonically increasing phase variables''~\cite[p.\,25]{Grindrod.2018}. This very question arises also, albeit in a more indistinct form, if GNW is formulated in terms of neuronal architecture: How does a ``piece of information selected for its salience or relevance to current
goals''~\cite[p.\,56]{DehaeneChangeuxNaccache.2011}, which is really just a state of some subset of the brain's neurons, relate to experience?
A proposal to this \ch{extent} is presented in Section~\ref{SecEFE}.

Whereas from a formal modelling perspective, there is some space for further development of the GNW model, it does seem to capture essential neuroscientific evidence in a simple and very plausible hypothesis: The global workspace. This idea might ultimately be combined with ideas of IIT or other models to give an explicit account of how the state of the global workspace relates to experience as we find it.

\subsection{Conscious Agent Networks}\label{ExCAN}
A model which is based on idealistic metaphysics is developed in~\cite{HoffmanPrakash.2014}.
The underlying idea is that what exists are interacting conscious agents, each of which has a fundamental capacity to perceive, decide and act, and that the interaction between these conscious agents seems to each as if there is an external outside world. For simplicty, in what follows, we explain a slight more general version of the model than presented in~\cite{HoffmanPrakash.2014}.

In order to explain a single conscious agent $C$, we first assume that there is a space $W$ which is external to the conscious agent. One may think of this as states of some ``world'' which the agent perceives, but in fact this space is constituted via interactions with other conscious agents, as explained below.
Given this space~$W$, a conscious agent is modelled as a five-tuple $C := \big( X, G, P, D, A \big)$, where $X$ and $G$ are spaces, and $P, D, A$ are maps,%
\footnote{In~\cite{HoffmanPrakash.2014}, the specification furthermore includes an integer $N$ which counts  perception-decision-action cycles
and hence acts as a type of internal ``psychological'' time, which however we simply replace by the usual parameter $t \in \I$.}
interpreted as follows:
\begin{itemize}[leftmargin=1.5em,topsep=0cm]
\itemD $X$ is a space which describes possible experiences of the conscious agent. Each element $x \in X$ represents a particular experience.
\itemD $G$ is a space which describes describes dispositions or intentions to act. Each element $g \in G$ corresponds to an action the agent has decided to carry out.
\itemD $P: W \rightarrow X $ is a map which describes the agent's ``process of perception''~\cite[p.\,6]{HoffmanPrakash.2014}. It specifies what the conscious agent experiences in response to the ``world'' being in a particular state $w \in W$.
\itemD $D: X \rightarrow G$ is a map which models how the experience of the agent determines its disposition for an action, i.e. ``the process of decision [in which] a conscious agent chooses what actions to take based on the conscious experiences it has.''~(ibid.).
\itemD $A: G \rightarrow W$ describes how the agent's disposition for an action ``is carried out'', i.e. how it affects the world: ``In the process of action, the conscious agent interacts with the world in light of the decision it has taken, and affects the state of the world''~(ibid.).
\end{itemize}

The structure of the spaces $W$ and $X$, as well as the definitions of the maps $P, G$ and $A$ are not fixed by the theory, but need to be chosen according to the application.\footnote{\label{CASimpl}In~\cite{HoffmanPrakash.2014}, some general assumptions are made: The spaces $W$, $X$ and $G$ are assumed to be measurable spaces and the maps $P$, $D$ and $A$ are chosen to be Markovian kernels, so that for every element of their domain, each map yields a probability distribution on their co-domain.}
Based on such a choice, the model specifies the dynamically possible trajectories
$ \big( x(t), g(t), w(t) \big)_{t \in \I}$ as those trajectories which satisfy
\begin{align*}
\big( x(t+1), g(t+1), w(t+1) \big) = \big( Pw(t), Dg(t), Aw(t) \big) \:,
\end{align*}
where $t$ is chosen as a discrete time parameter, i.e. $\I := \Z$.\smallskip

The central hypothesis of this theory is called ``conscious realism'': That ``[t]he world~$W$ consists entirely of conscious agents''~~\cite[p.\,7]{HoffmanPrakash.2014}. 
This hypothesis is implemented via networks of conscious agents.

In order to describe a network of $n$ conscious agents, we first assume that for every conscious agent, a space $X_i$ of possible experiences and and a space $G_i$ of dispositions to act is given, as well as a ``decision map'' $D_i$ as introduced above. The ``external world'' of the $i$th conscious agent is defined to be the product of the action spaces of all other conscious agents, i.e.
\[
	W_i := G_1 \times ... \times G_{i-1} \times G_{i+1} \times ... \times G_n \:.
\]
This choice is motivated by the idealistic idea that what exists are only experiences and dispositions to act, and that the dispositions to act of some agents determines the experience of others. I.e., the process of perception of the $i$th conscious agent is, in case of a network of conscious agents, given by a map
\[
	P_i : W_i \rightarrow X_i  \: .
\]
This allows us to define the dynamically possible trajectories of the network of conscious agents via
\[
x_i(t+1) =  P_i (g_1(t), ... \, , g_n(t) ) \quad \textrm{ and } \quad  g_i(t+1) = D_i x_i(t) \: .
\]
If $P_i$ is a partial function defined only for some $G_j \in W_i$, the $i$th agent is only able to perceive the dispositions to act of the corresponding other conscious agents. Various concrete proposals for how to choose $P_i$ are discussed in~\cite[p.\,7ff.]{HoffmanPrakash.2014}.

Due to the identification of the ``outside worlds''~$W_i$ of each conscious agent with the dispositions to act of others, the action map $A_i: G_i \rightarrow W_i$ is not
necessary to define the dynamics. In order to satisfy the definition of a conscious agent given above one may define it formally as the map which takes $g_i(t)$
to $w_i(t+1) = (g_1(t+1), ... \, , g_n(t+1) )$. In simple cases (e.g. involving two conscious agents~\cite{HoffmanPrakash.2014}) this definition can be flashed out in
terms of combinations of inverses of $D$ and $P$. In general, it may require $A_i$ to be time-dependent.\smallskip

In summary, the various objects the theory assumes in a particular application determine (possibly in a probabilistic manner) the dynamics of a network of conscious agents. The goal, then, is to specify plausible assumptions which allow \ch{us} to deduce formally that
``the perception of objects and space-time can emerge from such dynamics''~\cite[p.\,1]{HoffmanPrakash.2014}
and to ``explore [the model's] theoretical implications in the normal scientific manner to see if they comport well with existing data and theories, and make predictions that are novel, interesting and testable''~\cite[p.\,7]{HoffmanPrakash.2014}.

An early example of a result of this kind is given in~\cite[p.\,13ff.]{HoffmanPrakash.2014}. In a nutshell, it is shown that if the state spaces $X_i$ and $G_i$ are finite,
the dynamics of a network of two conscious agents can be described in terms of an object which bears some similarity to a quantum-mechanical wave function of a free particle.
\smallskip

From the perspective of models of consciousness as defined in Section~\ref{PhenGrMethod},
two crucial questions arise:
\begin{enumerate}[label=\alph*),topsep=0cm]
\item Whether the model would like to address aspects of experience which are non-collatable.
\item Whether the theory would (eventually or in principle) like to make predictions with respect to experiments which involve (reports of) conscious agents.
\end{enumerate}

An affirmative answer to the first question might be indicated by the remark that the ``qualia $X$ of a conscious agent $C$ are private, in the sense
that no other conscious agent $C_i$ can directly experience $X$''~\cite[p.\,14]{HoffmanPrakash.2014}.
If this is indeed the case, the mathematical structure of the spaces $X_i$ (and possibly also of $G_i$, if one holds that intentions to act are also experiences of some sort) could be defined based on a phenomenological analysis as explained in Section~\ref{PhenomenologicalGrounding}. 
This would, in particular, dismantle the objection that the ``definition of conscious agents could equally well-apply
to unconscious agents [so  that the] theory says nothing about consciousness''~\cite[p.\,14]{HoffmanPrakash.2014}.

More importantly, if the theory also answers affirmatively to the second question, the results of Section~\ref{PhenGrMethod} show that
a further mathematical structure is necessary to ensure that the model is empirically well-defined (Lemma~\ref{Necessary}).

\subsection{Expected Float Entropy Minimisation}\label{SecEFE}
One of the largest questions at present left open by the GNW model (Section~\ref{SecGNW}) is how the state of the global neuronal workspace, ultimately a collection of states of individual neurons, relates to experience.
Questions of this kind are addressed by the Expected Float Entropy (EFE) model developed in~\cite{Mason.2016}. In short, this is a proposal for how (probability distributions of) brain states determine relations among qualia.
\smallskip

In what follows, we review the definition of this model.
Every brain state is assumed to consist of individual elements, each of which can be in a particular state. We denote the set of all elements (``nodes'') by $S$ and the space of states of each node by $V$, and assume both are a finite set. A brain state is thus a map
\beq\label{ExRel14}
	s: S \rightarrow V \:.
\eeq
E.g., in a neural network, $S$ is the set of neurons and $V$ is the set of possible states of each neuron. If applied to the GNW model as outlined above, $S$ is the set $N_g$ of nodes and $V$ denotes the corresponding space of states. In~\cite{Mason.2016}, $s$ is called a ``data element'', but we will refer to $s$ simply as `state'.

Let $\Omega_{S,V}$ denote the space of all states. We assume that a probability distribution~$p$ is given over $\Omega_{S,V}$. The probability $p(s)$ can be interpreted as the probability of the brain being in state $s$.

A weighted relation on a set $S$ is a map $R: S \times S \rightarrow [0,1]$. Given a set of states with corresponding probability distribution,
the theory developed in~\cite{Mason.2016} allows \ch{one} to determine two weighted relations $R$ and $U$, where $R$ is a weighted relation on the set $S$ of nodes and where $U$ is a weighted relation on the possible states $V$ of each node. We will discuss the interpretation of $R$ and $U$ at the end of this example.

The theory determines both $U$ and $R$ as follows. For any state $s \in \Omega_{S,V}$, the composition $U( s (.) ,  s (.))$ is a relation on $S$, which we denote as $U \! \circ \!  s$. Define the float entropy $\fe$ and expected float entropy $\efe$ as 
\begin{align}\label{ExRel11}
	&\fe(R,U, s) = \log_2 \big| \big\{ \tilde{s} \in \Omega_{S,V} \, \big| \, d(R, U \! \circ \! \tilde{s} ) \leq d(R, U \! \circ \! s  \big\} \big|\\
	&\efe(R,U,p) = \sum_{s \in \Omega_{S,V} } p(s) \fe(R,U,s) \label{ExRel12}
\end{align}
where $s \in \Omega_{S,V}$, $d$ is a distance function on the weighted relations on $S$ and where $|A|$ denotes the cardinality of a set $A$.
The theory proposes ``that a system (such as the brain and its subregions) will define U and R (up to a certain resolution) under the requirement that the $\efe$ is minimized.'' I.e. $U$ and $R$ are defined via
\beq\label{ExRel13}
\efe(R,U,p) = \min_{\bar R, \bar U} \efe(\bar R, \bar U,p) \: ,
\eeq
where the minimum is taken over all relations $\bar R$ on $S$ and all relations $\bar U$ on $V$. (Existence or uniqueness of minimizers is not discussed in~\cite{Mason.2016}.)\smallskip

Concerning the interpretation of $R$ and $U$, the theory proposes that if ``a brain state is interpreted in the context of all these relations
(...), the brain state acquires meaning in the form of the relational content of the experience''.
If applied to the visual cortex, the theory aims to explain ``perceived relationships between different colours, the perceived relationships between different brightnesses, and the perceived relationships between different points in a person's field of view (giving geometry)''. 

These interpretations are supported by several examples in~\cite{Mason.2016}, where the theory is applied to pictures, so that $S$ is the set of all pixels and $V$ describes the possible colour values at each pixel, which implies that $U$ is a relation between colour values and $R$ is a relation between pixels.
The support for these interpretations becomes more difficult when the theory is being applied, e.g., to the visual cortex, for in this case $U$ is a relation on the states of the nodes where the nodes could be individual neurons or tuples of neurons in the visual cortex for example, and $R$ is a relation on the set of these nodes, making it somewhat unclear why in this case a relation $U$ might give an explanation of, e.g.,  why ``blue appears similar to turquoise but different to red''.

One can, however, simply take the theory at face value by accepting that the relata of $U$ and $R$, whichever mathematical form they take, {\em are} (describing) non-collatable aspects of experience and that $U$ and $R$ {\em are} (describing) the relations between them. Here, the non-collatability is essential for otherwise the identity of some collatable aspect of experience and elements of the set $V$ or $S$ would be questionable. In short, one may assume that the relations $R$ and $U$ correspond to the structure of the experience space $\E$ which describes experience.\smallskip 

Several interesting questions are raised by this model. First of all, we note that since the model aims to explain the relations between aspects of experience, it is fully compatible with a direct description of qualia as discussed in Section~\ref{Comparison} and does not aim for a description of qualia sensu stricto. This raises the question of whether this model is an alternative to, or rather a complement of, models which do intent to describe qualia sensu stricto, such as e.g. Integrated Information Theory.

One might conjecture that the relations among aspects of experience might in fact be nuanced enough to allow us to identify individual qualia by specification of the relations. In other words, that {\em all} orbits of the automorphism group~\eqref{AutGroup} are trivial.
Whether or not this is the case is a phenomenological question, which needs to be answered by a systematic account of the relations between qualia found in experience and is a priori to any model-building process (just like general properties of an explanandum have to be fixed prior to an explanation). However, since the EFE model actually specifies the relations between aspects of experience, one can also study which answer the model itself gives to this question. The upshot of this analysis, which is presented in the next paragraph, is that if the probability distribution $p$ is invariant with respect to a transformation (or permutation) of states, which is often the case, the model does in fact specify relations whose automorphism group has non-trivial orbits.

Consider a bijective transformation (permutation) of states $\sigma: \Omega_{S,V} \rightarrow \Omega_{S,V}$ which can be specified in terms of a bijective transformation $
\sigma_S: S \rightarrow S$ of nodes and in terms of a bijective transformation $\sigma_V: V \rightarrow V$ of node-states, i.e. $\sigma(s) := \sigma_V \circ s \circ \sigma_S$.
The probability distribution $p$ is {\em invariant} with respect to this transformation if $p = p \circ \sigma$, i.e. if the transformation maps states $s$ to states $\sigma(s)$ which have the same probability as the former, $p(\sigma(s)) = p(s)$. Defining the transformation of the relations~$U$ and~$R$ as
\begin{align}\label{ExRel15}
U'(.,.) := U(\sigma_V^{-1}(.), \sigma_V^{-1}(.)) \quad \textrm{ and } \quad R'(.,.) := R(\sigma_S(.), \sigma_S(.)) \:,
\end{align}
and using the fact that the metric $d$ is chosen as one of the $d_n$ metrics in~\cite[p.\,127]{Mason.2016}, i.e. involves summation over all elements of $S \times S$,
\eqref{ExRel11} yields that $\fe(R,U,s) = \fe(R',U',\sigma(s))$. Using the invariance of~$p$ and~\eqref{ExRel12}, this gives
\[
\efe(R,U,p) = \efe(R',U',p) \:.
\]
This implies that for any minimizer $R, U$ of~\eqref{ExRel13}, the pair $R',U'$ is a minimizer of~\eqref{ExRel13} as well. In other words, the theory only determines minimizers up to transformations~\eqref{ExRel15}. Assuming uniqueness of minimizers, this in turn implies that the minimizing pair $U,V$ satisfies%
\begin{align}\label{ExRel16}
U(.,.) = U(\sigma_V(.), \sigma_V(.)) \quad \textrm{ and } \quad R(.,.) = R(\sigma_S(.), \sigma_S(.)) \:,
\end{align}
so that $\sigma_V$ and $\sigma_S$ are relation-preserving bijections, i.e. non-trivial elements of the automorphism group of the spaces $(V,U)$ and $(S,R)$, respectively.\smallskip

Another interesting question is which part of the brain generates those relations between aspects of experience which we find in experience. This is, to a large \ch{extent}, a question which could be answered by simulations of the brain's neuronal network. If it turns out that these relations can be reproduced better by a distributed network, this model may actually be compatible, or even taken as support of, the Global Neuronal Workspace hypothesis. The underlying challenge here is, of course, to identify the weighted relations $R$ and $U$ between (states of) neurons with the manifold relations between aspects of experience. This identification may also hinge on how  the probability distributions $p(s)$, which is the only data which enters the definition of $R$ and $U$,  is interpreted when applied to the brain.\smallskip

We conclude that this theory is an interesting approach to the mind-matter relation which might complement more neuroscientific approaches such as the Global Neuronal Workspace model. Depending on whether a phenomenological analysis confirms that there are qualia which cannot be distinguished by mere reference to collatable relations, the model may or may not have to be extended in some form or the other to talk about the hard problem of consciousness.

\section{Conclusion \& Outlook}\label{Summary}

Consciousness is in the focus of research projects around the globe. Empirical as well as theoretical projects aim to investigate different aspects of experience, ranging from access consciousness or the unity of a conscious scene to phenomenal consciousness or the first-person-perspective~\cite{Seth.2007}. The starting point of this paper is the observation that if an aspect of experience is under investigation which cannot be identified over several experiencing subjects (which cannot be {\em collated}), special care is necessary. Any reference to such aspects of experience, be it in a theoretical account or when giving reports, is ambiguous and this ambiguity may lead to ill-defined models, erroneous empirical predictions and  misinterpretation of experimental data. A detailed summary of results is given in Section~\ref{sec:summary}.
\smallskip

In order to develop a well-defined scientific methodology which can be applied to all aspects of experience, we have used basic
{\em phenomenological \ch{axioms}} to specify how a formal representation of experience can be constructed. The result is a mathematical
space which represents some parts of experience (such as visual experiences or auditory experiences) {\em completely}, including both
the usual objects of investigation in cognitive neuroscience as well as qualia.
This formal representation of experience avoids the usual hard cut between parts of experience which represent a difficulty for the scientific methodology and parts which do not.  Both are interwoven in our formal representation, similarly \ch{to position and momentum being two aspects of a quantum state.}

We have shown that this mathematical representation of experience allows us to quantify the ambiguity involved in any reference to experience precisely. This is sufficient to avoid the problems mentioned above and yields a formal mathematical toolbox which can be applied in empirical or theoretical investigations of consciousness.\smallskip

In the second part of the paper, we have investigated how individual non-collatable aspects of experience (qualia ``sensu stricto''~\cite{IIT30}) can be studied scientifically. Since there is a fundamental explanatory gap, this question may be considered as equally relevant to the one addressed in the first step.

The main result of the second part of this paper is that formal models of consciousness can address individual non-collatable aspects of experience {\em if and only if} they carry a specific symmetry group related to the mathematical representation of experience explained above. Because of mathematical details of the action of this symmetry group, models of consciousness can be used to construct empirically well-defined theories of how individual aspects of experience relate to the physical domain {\em despite} the ambiguity inherent in any reference to the latter.\smallskip

The results of this paper constitute a grounding of the scientific study of consciousness which is an alternative to other groundings currently in use. It offers a thorough conceptual and mathematical framework in light of which existing models of consciousness can be interpreted and improved, and based on which new models can be constructed.

This constitutes a first step in developing a full-fledged conceptual and mathematical foundation for models of consciousness. Further work is necessary to investigate which mathematical structures are implied by other key characteristics of conscious experience, most notably the various connotations of subjectivity and intrinsicality, and to understand whether mathematical structure can be sufficient to account for any of them.\medskip

\Thanks {{\em{Acknowledgements:}}
I am grateful for the questions and comments received
during presentations of parts of this work at the the LPS Colloquium
of the Munich Center for Mathematical Philosophy, the Mathematical Institute of the University of Göttingen,
the Institute for Theoretical Physics of the University of Hanover, the Modelling Consciousness Workshop in Dorfgastein, 
the Models of Consciousness Conference in Oxford, the Online Seminar Progress and Visions in the Scientific Study of the Mind-Matter Relation and the Conceptual Foundations of Science Workshop in Tegernsee. Most of this work has been carried out while I was employed at the Institute for Theoretical Physics of the Leibniz University of Hanover, and I am very grateful for having had the opportunity to do so.\smallskip

\appendix
\section[Chalmers' Grounding of the Scientific Study of Consciousness]{Chalmers' Grounding of the Scientific Study of Consciousness}\label{ChalmersGrounding}
The most prominent grounding of the scientific study of consciousness has been developed by David Chalmers in~\cite{Chalmers.1996}.
Since it is the blueprint of the grounding proposes in Section~\ref{NewGrounding}, we review its essential definitions. Note, however, that the 
following outline of Chalmers' grounding is intended to highlight the relations among various constituents of his grounding
and is not intended to be of an introductory nature. A good and short introduction to this topic is~\cite[Ch.\,1]{Chalmers.2010}.

First, we note that Chalmers' definition of `physical domain' includes what is often called `material' or `physical' configurations, such as neurons or brain tissue,
as well as more fundamental physical notions such as ``mass, charge, and space-time''~\cite[p. 17]{Chalmers.2010} or ``atoms, electro­-magnetic fields, and so on''~\cite[p. 71]{Chalmers.1996}.
We thus define the term `{\em physical domain}'\label{PhysDomain}
to refer to all those phenomena which are currently considered to be the subject of a natural science (physics, chemistry, earth science, biology, etc.~\cite{Wikipedia.naturalscience}).
Chalmers assumes that:

\begin{enumerate}[label=(A\arabic*)]
\item \label{A1} ``The physical domain is causally closed.''~\cite[p.\,161]{Chalmers.1996}\\
``For every physical event, there is a physical sufficient cause.''~\cite[p.\,125]{Chalmers.1996}
\end{enumerate}

Central to Chalmers' grounding are the terms `function' and `structure'. ``Here `function' is not used in the narrow teleological sense of something that a system is designed to do but in the broader sense of any causal role in the production of behaviour that a system might perform''~\cite[p.\,6]{Chalmers.2010}. 
The term `structure' is used in a spatiotemporal sense. Together, they constitute, according to Chalmers, the notion of explanation which is used throughout contemporary science: ``One can argue that by the character of physical explanation, physical accounts explain {\em only} structure and function, where the relevant structures are spatiotemporal structures, and the relevant functions are causal roles in the production of a system's behavior.''~\cite[p.\,105f.]{Chalmers.2010} 

We denote this notion of explanation by \ref{E1}. Assuming some laws or theories relating to the physical domain as given ($=$ accepted by the scientific community by and large) and referring to them as `accepted theoretical notions', \ref{E1} might be put as follows:
\begin{enumerate}[label=(E\arabic*)]
\item \label{E1} An explanation specifies the function and structure of an explanandum
in terms of the the function and structure of accepted theoretical notions.
\end{enumerate}

The crucial aspect of Chalmers' grounding is to establish, in a consistent and explicit way, that there are phenomena, related to consciousness, to which no function or structure (as defined above) can be associated. It follows that these phenomena cannot be explained according to \ref{E1} and hence, if \ref{E1} indeed captures all notions of explanations which are used throughout contemporary science, that they cannot be explained by contemporary science. -- There is an ``explanatory gap''~\cite{Levine.1983,Chalmers.1996}. 
Chalmers refers to these phenomena as ``phenomenal concepts'', ``phenomenal qualities'' or ``qualia''~\cite{Chalmers.1996}.%
\footnote{In~\cite{Chalmers.2010}, he prefers to use the term `experience': ``Sometimes terms such as `phenomenal consciousness' and `qualia' are also used here, but I find it more natural to speak of `conscious experience' or simply `experience.'\,''~\cite[p.\,5]{Chalmers.2010}.}
We refer to these phenomena as `phenomenal aspects of consciousness':

\begin{enumerate}[label=(D\arabic*)]
\item \label{D1} \textit{Phenomenal aspects of consciousness} are those aspects of conscious experience which do not have a function or structure, where `function' and `structure' are as defined above.
\end{enumerate}

The key requirement for this definition of what is to be studied by a science of consciousness to make sense is to establish that there are aspects of experience which satisfy~\ref{D1}, i.e. which neither have a spatio-temporal structure nor a causal role in the production of behaviour. It is the second requirement with respect to which~\ref{A1} is crucial, for \ref{A1} can be utilized to argue that nothing non-physical can have a causal influence on the physical domain.
Therefore, all aspects of experience which do not have a spatio-temporal structure (e.g. in the Cartesian sense of being non-extended in space and space-time) automatically satisfy \ref{D1}.
We will not review the various arguments which aim to prove the existence of phenomenal aspects of consciousness at this point.

Put in terms of Definition~\ref{DefG}, what is to be studied in the scientific study of consciousness are, according to this grounding, phenomenal aspects of consciousness and their relation to the physical domain. 
Since these are, by definition, not accessible to the usual scientific methodology, Chalmers proposes that the task of a science of consciousness is to find what he calls ``psychophysical laws''~\cite[p.\,127]{Chalmers.1996} which relate the physical domain to phenomenal aspects of consciousness. Due to Assumption~\ref{A1} and an underlying stance on the nature of causality
``[t]hese laws will not interfere with physical laws; physical laws already form a closed system. Instead, they will be {\em supervenience laws}, telling us how experience [= phenomenal aspects of consciousness] arises from physical processes''~\cite[p.\,127]{Chalmers.1996}.
In combination with \ref{E1}, this implicitly points at the major parts of the methodology to be used according to this grounding.

Chalmers' grounding raises several questions related to the definition and ontological status of causality, to the validity of Assumption~\ref{A1}, to the nature of experiments in his grounding and to the validity of the subsumed notion of explanation, which we discuss in Appendix~\ref{APIssuesChalmers}. 
The upshot is that there are severe conceptual problems which make it questionable whether a scientific research program based on this grounding can be carried out at all.

Furthermore, any scientific approach based on this grounding faces the question of which mathematical structure one is to use in order to describe phenomenal aspects of consciousness when formulating ``psychophysical laws''~\cite[p.\,127]{Chalmers.1996}. Whereas the physical domain comes with a clear-cut mathematical structure, Chalmers' grounding merely asserts that the phenomenal aspects form a set and offers no systematic way of tying additional mathematical structure to the phenomenology of experience.

This  strongly suggest the construction of other groundings of the scientific study of consciousness. 
In Section~\ref{NewGrounding}, we have introduced a possible alternative which avoids the above-mentioned problems. Whereas this grounding breaks with several of Chalmers' main ideas, it retains the key idea of addressing an explanatory gap with mathematical tools.

\section{Conceptual Problems of Chalmers' Grounding}\label{APIssuesChalmers}\addtocontents{toc}{\setcounter{tocdepth}{-10}}
In this appendix, we briefly discuss several conceptual issues of Chalmers' grounding. These issues are not motivated by metaphysical considerations and are not intended to have metaphysical implications; they simply arise if one wishes to carry out a scientific investigation of consciousness based on Chalmers' grounding.
Problems~\ref{AppendixClosure} and~\ref{AppendixExperiments} are most crucial and might make it impossible to apply the grounding.

The abbreviations used below have been introduced in Appendix~\ref{ChalmersGrounding}.
For reasons explained in Appendix~\ref{SubSecACaus}, we assume that Assumption~\ref{A1} is intended to express the fact that ``physical laws already form a closed system''~\cite[p.\,127]{Chalmers.1996}.

\subsection{Closure of the Physical}\label{AppendixClosure}
Much has been written about Assumption~\ref{A1} both by David Chalmers himself (e.g.~\cite[Ch.\,5]{Chalmers.1996} or~\cite[Ch.\,8 and\,9]{Chalmers.2010}) and by others (e.g. \cite{Elitzur.2009} or~\cite{Bishop.2005}).
As noted in Section~\ref{ChalmersGrounding}, this assumption is crucial for Chalmers' grounding in order to establish that there are aspects of experience which satisfy~\ref{D1}.

To date, there is no valid argument which shows that Assumption~\ref{A1} is wrong, i.e. that the physical laws of nature cannot ``form a closed system''. On the other hand, there also is no valid argument that shows that Assumption~\ref{A1} is right, i.e. that the physical laws of nature must form a closed system.\footnote{Note that no reasons are given in either~\cite{Chalmers.1996} or~\cite{Chalmers.2010}
for why Assumption~\ref{A1} should hold true.}
This assumption also cannot be backed by analysing opinions or strategies of working physicists, for most physicists are prepared to accept, or even try to find, modifications of the known laws of physics due to yet unknown phenomena (e.g. related to dark matter, to quantum gravity or to dynamical collapse theories, to name just a few). They do not assume that the known physical laws form a closed system. ``Physics itself does not imply its own causal closure nor is there any proof within physics of its own completeness, so CoP [causal closure of physics] must be a metaphysical principle''~\cite[p.\,45]{Bishop.2005}.

Based on this state of affairs, one might think that both Assumption~\ref{A1} as well as its opposite should be compatible with a scientific approach to consciousness. However, this is not the case, as the following remark shows. Despite the fact that the physical laws of nature may form a closed system, it seems that
Assumption~\ref{A1} is incompatible with a scientific approach to investigate consciousness because it violates a necessary condition for the possibility of the latter.

\begin{Remark}\em \label{NoClosureOfPhys}
The phenomenological grounding developed in Section~\ref{NewGrounding} allows \ch{one} to construct models of consciousness which postulate the physical as closed just as well as models which do not postulate the physical as closed (several examples of both are given in Section~\ref{Examples}). However, it seems that in both Chalmers' and the phenomenological grounding of the scientific study of consciousness, it does {\em not} make sense to {\em assume} the closure of the physical because it violates a necessary condition of the possibility of the scientific study of consciousness itself. 

The goal of this remark is to explain in detail why this is so. To this end, we use the symbol~$\mathcal Q$ to denote that which is to be studied according to the grounding at hand: In the case of Chalmers' grounding (CG), $\mathcal Q$ refers to qualia as defined in~\ref{D1} in Appendix~\ref{ChalmersGrounding}, whereas in the case of the phenomenological grounding (PG), $\mathcal Q$ refers to qualia as defined in Definition~\ref{DefQualia}. In both cases, $\mathcal Q$ thus refers to aspects of conscious experience.

The above claim rests on two premises. First, that a {\em scientific} study of consciousness is possible only if scientists can {\em communicate} about $\mathcal Q$ at least {\em to some \ch{extent}}. E.g., they need to be able to agree on $\mathcal Q$'s definition and existence, need to be able to communicate certain general properties of $\mathcal Q$ (such as Phenomenological Axioms~\ref{PFcollatable},~\ref{PFRecognize} or~\ref{PFRel} in the case of PG) or need to be able to record and exchange data related to~$\mathcal Q$. This is the necessary condition for the possibility of the scientific study of consciousness referred to above, which we abbreviate by NC.

The second premise is that communication is always mediated via {\em communication channels}~$\mathcal C$ which are elements of the physical domain. To give some examples, consider verbal communication, which is mediated via sound waves, digital communication, which is mediated via electromagnetic signals, or printed texts, where communication is mediated via arrangements of molecules and electromagnetic fields.

Due to the second premise, an assumption concerning the closure of the physical (ACoP) has something to say about communication channels and therefore also about communication itself. If, in the grounding at hand, ACoP is fleshed out in such a way that it restricts the
relation between~$\mathcal Q$ and communication channels~$\mathcal C$ to such an \ch{extent} that communication about~$\mathcal Q$ is impossible, the above claim holds: By the first premise, this implies a violation of a necessary condition of the possibility of a scientific study of consciousness.

Clearly, whether or not this is the case depends on what one takes to constitute `communication' and which conditions one posits as necessary for something to count as `communication about $\mathcal Q$'. To find proper answers to these questions is of course the goal and task of various parts of philosophy. However, by restricting to a very simple situation, we may hope to work with a necessary requirement for `communication about $\mathcal Q$' to be possible which is acceptable independently of which notion of communication one prefers.

The simple situation which we consider is the prototypical scenario of the mathematical theory of communication.\footnote{This is 
the original (and arguably more adequate~\cite{Floridi.2017}) name for `information theory'~\cite{Shannon.1948}.}
I.e., we consider a situation where one experiencing subject~$\mathcal S_1$ (the `sender') formulates a message $m_1$ which expresses
some properties of her experience of~$\mathcal Q$, such as which particular phenomenal quality she has experienced (in the case of CG) or whether two qualia are similar (in the case of PG).%
\footnote{If this is impossible, i.e. if the assumptions of a grounding are such that an experiencing subject cannot formulate a message which expresses some properties of her experience of~$\mathcal Q$, this grounding violates the necessary condition NC as claimed. This may be the case for CG,
cf.~\cite[Ch.\,9]{Chalmers.2010}.}
Subsequently, this message is being transferred via a communication channel~$\mathcal C$ to another experiencing subject~$\mathcal S_2$ (the `receiver'), who after decoding the channel's signals obtains a message $m_2$. We abbreviate this scenario by MTCp (`p' for `prototypical').

We denote properties of $\mathcal S_1$'s experience of $\mathcal Q$ by~$q$ and states of the communication channel~$\mathcal C$ by~$c$. In what follows, we consider functional dependencies between the quantities $q$, $c$, $m_1$ and $m_2$. 
In order to define what constitutes a functional dependency both mathematically and conceptually, we refer to the groundings under consideration:

Both CG and PG's specification of the task of the scientific study of consciousness includes the formulation of laws or theories concerning the relation of~$\mathcal Q$ with the physical domain. Given enough further specifications (such as a model of the communication channel or more comprehensive physical laws), these laws or
 theories should be applicable to the MTCp setup. I.e., we may assume that both PG and CG allow \ch{one} to construct (or even to deduce) mathematical models of the MTCp setup. The details of any such model of course depend on various factors, most importantly on which psychophysical laws (CG) or models of consciousness (PG) one considers. All that matters at this point is that given any such model, we may identify functional relationships between the quantities $q$, $c$, $m_1$ and $m_2$:
\bei
\item[(F1)] A quantity $a \in \{q,c,m_1,m_2\}$ is {\em functionally dependent} on a quantity $b \in \{q,c,m_1,m_2\}$ according to some model of the MTCp setup
iff according to this model, $a$ is a non-constant function of $b$.%
\footnote{Here, by `constant function' we simply refer to functions which are formally dependent on $b$ but whose value remains the same independently of which value~$b$ takes. E.g., $f(x,y) := x$ is a constant function of $y$.}
\eni

The reasons for focussing on functional dependency in order to argue for the main claim of this remark are threefold.
The first reason is that in both CG and PG, ACoP implies a restriction of the functional dependencies which may hold between the quantities $q$, $c$, $m_1$ and $m_2$. 
Consider first CG. Here, the various formulations of ACoP differ slightly depending on whether they utilize a notion of causality or not. However, it seems fair to say that they all intend to express the central claim that ``physical laws already form a closed system''~\cite[p.\,127]{Chalmers.1996}. Together with the second premise introduced above, this implies that in any model of the MTCp setup based on CG, $c$ cannot functionally depend on~$q$.
In PG, ACoP implies that any state $c$ of the communication channel is determined completely by the dynamics of the physical theory~$T_P$, which does not include~$\mathcal Q$. Therefore, as is the case for CG, in PG ACoP also implies that $c$ cannot functionally depend on~$q$:
\bei
\item[(A2)] In both Chalmers' grounding (CG) and the phenomenological grounding (PG), the assumption of the closure of the physical (ACoP) implies that the states~$c$  of communication channels {\em cannot} be functionally dependent on~$q$.%
\footnote{\label{FunctionalDep1}We emphasize again that the notion of `functional dependence' is defined by the respective grounding under consideration. Thus it has a somewhat nomological flavour and does not express, e.g., simple covariation. The fact that both groundings contain notions of functional dependence is what allows the present argument to be stated in a comparably concise form.}
\eni
\noindent The second reason is that functional dependency also seems to allow \ch{us} to formulate a fundamental necessary condition for `communication about~$\mathcal Q$' to be possible:
\bei
\item[(C1)] A necessary condition for {\em communication between $\mathcal S_1$ and $\mathcal S_2$ about $\mathcal Q$} is that $m_2$ may depend functionally on $q$.
\eni
The third reason, finally, is that the MTCp is intended to express functional relationships in the first place. In particular, it can be taken to imply by definition that $m_2$ is functionally dependent only on $c$ and acquires additional functional dependencies only via $c$'s functional dependencies.

This concludes the reasoning: A necessary condition of communication about~$\mathcal Q$ in the MTCp setup is that~$m_2$ is functionally dependent on~$q$. By definition of the MTCp setup it can only be functionally dependent on~$q$ via $c$. CG and PG's ACoP however imply that~$c$ cannot be functionally dependent on~$q$. Therefore, a necessary condition of communication about~$\mathcal Q$ is violated, which by the first premise above is a violation of a necessary condition for the possibility of a {\em scientific} study of consciousness.\smallskip

Clearly, this reasoning does not yet constitute a formal argument. Several of its suppositions have to be checked carefully for hidden assumptions, which goes beyond the scope of this remark.\footnote{To give one example: As explained in Footnote~\ref{FunctionalDep1}, this argument rests on the notion of functional dependency 
contained in CG and PG in virtue of psychophysical laws or models of consciousness. In using these, we have avoided the difficult question of what a functional dependency actually expresses (i.e. how it is supposed to be defined and interpreted). E.g., when considering $q$, $c$, $m_1$ and $m_2$ as variables, which sort of possible words do they describe? Logically possible worlds, conceivable worlds, some sort of nomologically possible worlds? In what way can the assumptions of a grounding
restrict these possible worlds and what effect does this have on functional relationships?}
Nevertheless, it is of importance both with respect to Chalmers' grounding (where it raises a thorough problem) and with respect to the phenomenological grounding (where it is a basis for potential empirical predictions).\smallskip

We close this remark by pointing out that arguments which try to prove that the closure of the physical {\em cannot} hold in light of empirical facts about our experience (most notably written or verbal statements which express some fact about conscious experiences, e.g. bafflement about why consciousness exists~\cite{Elitzur.2009}) do not seem to be valid. The problem is simply that we may appear to be expressing facts about our conscious experience while in fact we are not. Similarly, we may appear to be communicating about consciousness while in fact we are not. This is the basis for Chalmers' efforts to develop a theoretical account of how judgements or statements about consciousness can be accounted for despite the closure of the physical, cf.~\cite[Ch.\,5]{Chalmers.1996} and~\cite[Ch.\,8 and\,9]{Chalmers.2010}.

In contrast, the claim proposed in this remark simply represents a transcendental argument: Independently of whether reality satisfies the closure of the physical or not, it does not make sense to engage in a {\em scientific} study of consciousness if one postulates the physical as closed, because the latter violates a necessary condition of the possibility of the former.
\QEDrem
\end{Remark}

\subsection{Experiments}\label{AppendixExperiments}
An issue also arises with respect to experiments if one postulates that ``physical laws already form a closed system''~\cite[p.\,17]{Chalmers.2010}: Almost all experiments one might wish to perform are rendered meaningless. The reason is simply that most experimental data (fMRI scans, EEG signals, verbal reports, etc.) is stored on physical devices (hard drives, paper, sound waves, etc.) and hence subject to physical laws. If these are postulated to ``form a closed system'' it follows that the experimental data must be determined by these physical laws alone, independently of which ``psychophysical law''~\cite[p.\,127]{Chalmers.1996} correctly describes how phenomenal properties depend on physical properties.\smallskip

To see this in more detail, let us assume that two different psychophysical laws~$\L$ and~$\L'$ have been proposed. 
The idea behind Chalmers' and in fact any conception of the {\em scientific} study of consciousness
is that experiments have to be carried out in order to evaluate which of the proposals better
describes reality.
Accordingly, assume  that an experiment has been designed and carried out which purports 
to answer this question, e.g. by checking predictions based on the laws~$\L$ and~$\L'$. 
Finally, denote by~$d$ the dataset produced by this experiment.

The term `data' is applicable to any ``putative fact regarding some difference or lack of uniformity within some context''~\cite{Floridi.2017}, so that one might consider the case where~$d$ actually consists of non-physical quantities, e.g. of differences in one's own experience. 
However, as soon as the data is stored or processed as usual, e.g. on a hard drive in order to perform statistical analysis, the differences in question have been transformed into ``difference or lack of uniformity'' of physical quantities. Since almost all experiments, even when dealing with verbal reports or similar indications of conscious experience, perform some sort of statistical analysis, it seems that in almost all experiments, $d$ eventually is a physical data set in this sense:%
\footnote{If one assumes that communication between two experiencing subjects is mediated via communication channels that are part of the physical domain (cf. Remark~\ref{NoClosureOfPhys}), it follows that every scientifically meaningful data needs to be transformed into physical data at some point.} It is `stored via' physical quantities.

If one assumes that ``physical laws already form a closed system''~\cite[p.\,17]{Chalmers.2010}, it follows that all physical quantities, as well the differences or lack of uniformity they exhibit, are determined by the laws of physics alone. Applied to the physical quantities on which~$d$ is stored,
this statement literally says that the data~$d$ is determined by the laws of physics alone.
Put differently, due to the fact that the experimental data~$d$ is stored on a physical device, closure of the physical implies that the data~$d$ is completely independent of whether~$\L$ or~$\L'$ or some completely different psychophysical law best describes how experience arises from physical processes.\smallskip

Thus, in summary, the closure of the physical implies that whatever experiment one performs in order to evaluate psychophysical laws, if it yields data that is stored on physical devices, the result of the experiment is independent of how experience actually arises from physical processes, i.e. independent of that which it seeks to study.

This conclusion holds true even if we concede that every experiencing subject might interpret the physical dataset~$d$ in terms of his/her own experience, so as to give meaning to this set in a way that a philosophical zombie might not,
simply because if~$d$ is independent of which law~$\E$ best describes how phenomenal properties arise from physics, the meaning a scientist gives to~$d$ will generally be too.%
\footnote{One may be able to avoid this last conclusion by insisting that the meaning attributed to~$d$ by any experiencing subject is dependent on the law~$\E$ itself and if one furthermore argues that
a conclusion about which law~$\E$ best fits nature can be deduced from the meaning of~$d$, despite~$d$ itself being determined independently of the former. At the present stage it seems quite unclear how such an deduction might work, let alone what role an experiment might play in this deduction in the first place.}

\subsection{Subsumed Notion of Explanation}\label{ChalmersExpl}
Chalmers' grounding builds on, axiomatizes and extends the notion of an explanatory gap that has been introduced by Joseph Levine in~\cite{Levine.1983}.
To this end, Chalmers claims that a specific account of explanation covers all notions of explanation that are used throughout natural science: An account in terms of function and structure, cf.~\ref{E1} in Appendix~\ref{ChalmersGrounding}. He subsequently shows that there are aspects of experience which do not have any of these two properties, so cannot be explained in terms of natural science as usual.
The gist of his grounding is that they may be addressed by a ``new sort of explanation''~\cite[p.\,121]{Chalmers.1996} which consists of 
 ``new fundamental laws (...) specifying how phenomenal (or protophenomenal) properties depend on physical properties''~\cite[p.\,127]{Chalmers.1996}.
 
The question of how scientific explanation is to be defined has occupied many philosophers throughout the 20th century~\cite{Woodward.2017}.
To find a definition which is general enough to capture the various explanations in science, yet specific enough to exclude scenarios which are clearly not cases of scientific explanation turns out to be a very difficult task. 
Even basic questions such as whether or not causality is to feature in the definition of explanation (and if yes, which definition of causality), are still largely debated:
``There is considerable disagreement among philosophers about whether all explanations in science and in ordinary life are causal and also disagreement about what the distinction (if any) between causal and non-causal explanations consists in.''~\cite{Woodward.2017}.

This sheds some doubt on Chalmers' notion of explanation, and the question arises whether~\ref{E1} really covers all, or even the most essential, uses of explanation throughout sciences. This is particularly so with respect to physics, whose notion of explanation seems to be a lot more formal than suggested by the terms `function' and `structure' as defined here. E.g., physics does seem to provide notions of explanation which can be applied to general dynamical quantities, whether they describe changes in the behaviour of a system%
\footnote{Recall that the term function refers to ``any causal role in the production of behavior that a system might perform''~\cite[p.\,6]{Chalmers.2010}.
One could interpret this as referring to ``any {\em change} in the behavior of a system'' (cf. Appendix~\ref{SubSecACaus}). This could, in turn, be taken to mean ``any change in the dynamical properties of a system'',
which would change the meaning of the claim that ``physical accounts explain {\em only} structure and function''~\cite[p.\,105f.]{Chalmers.2010} 
to the following:

``Any account given in purely physical terms will suffer from the same
problem. It will ultimately be given in terms of the structural and dynamical
properties of physical processes, and no matter how sophisticated such an
account is, it will yield only more structure and dynamics. While this is
enough to handle most natural phenomena, the problem of consciousness
goes beyond any problem about the explanation of structure and function [sic],
so a new sort of explanation is needed.''~\cite[p.\,121]{Chalmers.1996}

However, most or even all aspects of consciousness are dynamical in nature, which implies that the set of phenomenal aspects of consciousness (Definition~\ref{D1}) is, given this redefinition of the term `function',  either empty or trivial.
Put differently, with this redefinition the grounding implies that all or almost all of conscious experience can be addressed by an ``account given in purely physical terms''. What is left out are only non-dynamical aspects of experience (if there are such aspects at all).
}
or changes of a more general sort. (Chalmers might even reluctantly agree to this last observation when claiming that ``throughout the higher-level sciences, reductive explanation works in just this [\ref{E1}] way''~\cite[p. 7]{Chalmers.2010}, thus, in this quote, avoiding the claim that \ref{E1} also applies to lower-level sciences, such as (presumably) physics.)

This is a problem because the legitimacy of proposing ``new fundamental laws'' which describe how phenomenal aspects of experience depend on physical properties \cite[p.\,127]{Chalmers.1996}, as compared to a reductive explanation in terms of physical accounts, is granted, in Chalmers' grounding, by the existence of an explanatory gap between phenomenal aspects and contemporary scientific explanation. If scientific explanation is more powerful than Chalmers assumes, the justification of this explanatory gap breaks down and it becomes questionable whether this explanatory gap actually exists. 
``[A]n explanatory gap (...) cannot be made more precise than the notion of explanation itself''~\cite[p.\,358]{Levine.1983}.

\subsection{Causality}\label{SubSecACaus}
Finally, the question arises of what exactly one should take to constitute causality
when applying Chalmers' grounding. This is so because Assumption~\ref{A1} as well as Definitions~\ref{E1} and~\ref{D1} all relate to causality in an essential way (the latter via the definition of the term `function', cf. Appendix~\ref{ChalmersGrounding}).

This question is widely debated both in physics and in the philosophy of causation~\cite{Schaffer.2016}. It seems fair to say that consensus is missing
on basically all aspects of a definition of causality, including basic questions such as which relata a causal relation is to refer to and how, given a choice of relata, causality is defined. 
Whereas this multitude of possible notions of causality may not matter much if one is concerned with philosophical investigations based on Chalmers' grounding (one may just restrict to analyses that apply to every notion of causality), it does matter if one wishes to apply the grounding.
In particular, if one wishes to model, let alone to identify, phenomenal aspects of experience, one does need to know what
exactly the Definition~\ref{D1} amounts to. Since the term `function' used in that definition refers exclusively to causality, the defining property of phenomenal aspects depends on what one takes causality to be.

Connected to questions of how to define causality is the question of the ontological status of causality. Does some definition of causality pertain to ``reality'' or the universe? (In physicists' terms: Is causality ``fundamental''? Does the universe ``obey'' one particular definition of causality?) Or is causality rather a tool which can be utilized (by humans, animals, etc. or by information processing systems in general) to describe some parts of reality well to some \ch{extent}?\footnote{E.g.,~\cite{Pearl.2009} holds that ``[i]f you wish to include the entire universe in the model, causality disappears because interventions disappear -- the manipulator and the manipulated lose their distinction.
However, scientists rarely consider the entirety of the universe as an object of investigation. In most cases the scientist carves a piece from the universe and proclaims that piece {\em in} -- namely, the {\em focus} of investigation. The rest of the universe is then considered {\em out} or {\em background} and is summarized by what we call {\em boundary conditions}. This choice of {\em ins} and {\em outs} creates asymmetry in the way we look at things, and it is this asymmetry that permits us to talk about `outside intervention' and hence about causality and cause-effect directionality.''~\cite[p.\,419f.]{Pearl.2009}
``What we conclude (...) is that physicists talk, write, and think one way and formulate physics in another.''~\cite[p.\,407]{Pearl.2009}}

Chalmers' grounding is strongly dependent on which answer one gives to this question.
E.g., it determines which sort of ``influence'' of the phenomenal domain on the physical domain is compatible with the definitions of the grounding, or which type of condition the definition of phenomenal aspects of consciousness constitutes.
One may ignore this problem as long as one applies the grounding to theories of the physical domain which incorporate some notion of causality, such as, arguably, abstract neural networks. However, if one wishes to apply the grounding to fundamental physical theories, whose laws do not refer to, or come equipped with, any notion of causality, this question cannot be ignored.

These issues can be avoided completely if one takes the various uses of the term ``causality'' in Chalmers' grounding to jointly mean that the physical domain is not changed in any way by phenomenal aspects of consciousness, i.e., that the various uses of causality simply amount to ensuring that the ``physical laws already form a closed system''~\cite[p.\,127]{Chalmers.1996}. This seems to be the actual intention of the author in~\cite{Chalmers.1996} and~\cite{Chalmers.2010}, which is why
we have fixed this interpretation in the beginning of this appendix.

\addtocontents{toc}{\setcounter{tocdepth}{3}} 


\newcommand{\etalchar}[1]{$^{#1}$}

\end{document}